\documentclass[11pt,fleqn]{article}

\usepackage{amsmath,amssymb,amsthm,enumerate}

\newcommand{\p}{\partial}
\newcommand{\const}{\mathop{\rm const}\nolimits}
\newcommand{\rank}{\mathop{\rm rank}\nolimits}
\newcommand{\pr}{\mathop{\rm pr}\nolimits}

\newcommand{\CV}{\mathop{\rm CV}\nolimits}
\newcommand{\CL}{\mathop{\rm CL}\nolimits}
\newcommand{\Ch}{\mathop{\rm Ch}\nolimits}
\newcommand{\Eop}{{\sf E}}
\newcommand{\Fder}{\mathop{\sf D}\nolimits}
\newcommand{\ord}{\mathop{\rm ord}\nolimits}

\newcommand{\todo}[1][\null]{\ensuremath{\clubsuit}}

\newtheorem{theorem}{Theorem}
\newtheorem{lemma}{Lemma}
\newtheorem{corollary}{Corollary}
\newtheorem{proposition}{Proposition}
{\theoremstyle{definition} \newtheorem{definition}{Definition}

\newtheorem{note}{Note}

\begin{document}

\par\noindent {\LARGE\bf
Potential Conservation Laws \par}

{\vspace{4mm}\par\noindent \it 
Michael KUNZINGER~$^\dag$ and Roman O. POPOVYCH~$^\ddag$
\par\vspace{2mm}\par}

{\vspace{2mm}\par\it 
\noindent $^{\dag,\ddag}$Fakult\"at f\"ur Mathematik, Universit\"at Wien, Nordbergstra{\ss}e 15, A-1090 Wien, Austria
\par}

{\vspace{2mm}\par\noindent \it
$^{\ddag}$~Institute of Mathematics of NAS of Ukraine, 3 Tereshchenkivska Str., Kyiv-4, Ukraine
 \par}

{\vspace{2mm}\par\noindent $\phantom{^{\dag,\ddag}}$\rm E-mail: \it  $^\dag$michael.kunzinger@univie.ac.at, $^\ddag$rop@imath.kiev.ua
 \par}

{\vspace{5mm}\par\noindent\hspace*{8mm}\parbox{140mm}{\small
We prove that potential conservation laws have characteristics depending only 
on local variables if and only if they are induced by local conservation laws. 
Therefore, characteristics of pure potential conservation laws have to essentially depend on potential variables. 
This statement provides a significant generalization of results of the recent paper by Bluman, Cheviakov and Ivanova 
[{\it J. Math. Phys.}, 2006, V.47, 113505].
Moreover, we present extensions to gauged potential systems, Abelian and general coverings 
and general foliated systems of differential equations.
An example illustrating possible applications of proved statements is considered. 
A special version of the Hadamard lemma for fiber bundles and the notions of weighted jet spaces are proposed 
as new tools for the investigation of potential conservation laws. 
}\par\vspace{4mm}}



\section{Introduction}

In a recent paper by Bluman, Cheviakov and Ivanova~\cite{Bluman&Cheviakov&Ivanova2006} 
a remarkable result on potential conservation laws was obtained. 
Namely, it was shown that for an arbitrary system of differential equations 
a conservation law of a potential system with a characteristic 
which depends only on the independent variables is induced by a local conservation law of the initial system. 
It appears that this statement was deduced after an in-depth investigation of important examples on potential symmetries
which were considered ibid. This approach seems natural since,  
according to the famous Russian mathematician Vladimir Arnold, mathematics is an inductive and experimental science. 
In the present paper we show that this theorem admits a significant generalization and that, moreover, a converse statement is true as well. 
The possibility of deriving this result is suggested by recalling the rule of transforming conservation laws 
under point transformations between systems of differential equations~\cite{Popovych&Ivanova2004CLsOfNDCEs,Popovych&Kunzinger&Ivanova2008}. 
The application of a hodograph type transformation to a characteristic 
which exclusively depends on the independent variables may result in a characteristic including dependent variables. 
Generally, characteristics of induced conservation laws of potential systems can depend on derivatives of 
unknown functions of the initial system, and systems of other kinds related to standard potential systems 
(systems determining Abelian or general coverings, gauged potential systems, general foliated systems) 
can be investigated in the same framework.  

More precisely, we rigorously prove a number of statements on this subject 
(Proposition~\ref{PropositionOnInducedCLSOfFoliatedSystems} and 
Theorems~\ref{TheoremOnInducingExtendedCharsOfFoliatedSystems}--\ref{TheoremOnCLsOfMultiDPotSystems}), 
which can be summed up as follows:

\begin{theorem}\label{TheoremUnitedOnCLsOfDifferentPotSystems}
The following statements on a conservation law of a two-dimensional potential system 
(resp.\ a system determining an Abelian covering, resp.\ a multi-dimensional standard potential system without gauges)
are equivalent if the corresponding initial system is totally nondegenerate:

1) the conservation law is induced by a conservation law of the initial system;

2) it contains a conserved vector which does not depend on potentials;

3) some of its extended characteristics are induced by characteristics of the initial system;

4) it possesses a characteristic not depending on potentials. 

\noindent
The equivalence of the first three statements is also true for conservation laws of general foliated systems, 
including multi-dimensional gauged potential systems and covering systems.  
\end{theorem}

Further results on conservation laws of weakly gauged potential systems (Theorem~\ref{TheoremOnCLsOfMultiDWeaklyGaugedPotSystems})
and general covering systems (Theorem~\ref{TheoremOnCLsOfGeneralCoverings}) are established as well. 

Theorem~\ref{TheoremUnitedOnCLsOfDifferentPotSystems} allows us to formulate 
a criterion (Proposition~\ref{PropositionOnEssentialChars}) on purely potential conservation laws in terms of characteristics. 
Namely, \emph{a conservation law of a system determining an Abelian covering (resp.\ a potential system in the two-dimensional case) 
is not induced by a conservation law of the corresponding initial system if and only if 
it is associated with a completely reduced characteristic depending on potentials}.
Here, a characteristic of a system of differential equations is called completely reduced 
if it does not depend on the derivatives of the unknown functions, which are assumed to be constrained to the solution set of the system. 
In particular, any completely reduced characteristic of a system determining an Abelian covering does not depend on
the derivatives of potentials of orders greater than 0 since they are constrained 
due to differential consequences of the potential part of the system.
Any conservation law possesses a completely reduced characteristic since 
expressing the constrained variables via the unconstrained ones in a characteristic results in an equivalent characteristic. 

Our paper is organized as follows: 
Some basic notions and results 
on conservation laws are collected in Section~\ref{SectionOnBasicDefsAndStatementsOnCLs} for the reader's convenience.
Results on characteristics of conservation laws are singled out in Section~\ref{SectionOnCharacteristicsOfConsLaws} 
due to their particular importance for the paper. 
The exposition in these two sections follows, in general, the well-known textbook by Olver~\cite{Olver1993} 
while at the same time taking into account \cite{Popovych&Ivanova2004CLsOfNDCEs,Popovych&Kunzinger&Ivanova2008,Zharinov1986}.
Two versions of the Hadamard lemma for fiber bundles, which play a crucial role for our further considerations, 
are formulated and proved in Section~\ref{SectionOnHadamardLemmaForFiberBundles}. 
Then we successively study conservation laws of 
general foliated systems (Section~\ref{SectionOnFoliatedSystemsOfDifferentialEquations}), 
potential systems with two independent variables (Section~\ref{SectionOn2DimCase}), 
systems determining Abelian coverings (Section~\ref{SectionOnAbelianCoverings}), 
standard and gauged potential systems in the multi-dimensional case (Section~\ref{SectionOnStandardPotsInMultiDimCase}) 
and general covering systems (Section~\ref{SectionOnGeneralCoverings}). 
The criterion for purely potential conservation laws is formulated in Section~\ref{SectionOnCriterionForPurelyPotentialCLs}.
Possible applications of the obtained results are illustrated by an example in the final section.

\section{Basic properties of conservation laws}\label{SectionOnBasicDefsAndStatementsOnCLs}

Let~$\mathcal L$ be a system~$L(x,u_{(\rho)})=0$ of $l$ differential equations $L^1=0$, \ldots, $L^l=0$
for $m$ unknown functions $u=(u^1,\ldots,u^m)$
of $n$ independent variables $x=(x_1,\ldots,x_n)$.
Here $u_{(\rho)}$ denotes the set of all the derivatives of the functions $u$ with respect to $x$
of order no greater than~$\rho$, including $u$ as the derivative of order zero.
It is always assumed that the set of differential equations forming the system under consideration 
canonically represents this system and is minimal. 
The minimality of a set of equations means that no equation from this set is a differential consequence of the other equations.
By $L_{(k)}$ we will always denote a maximal set of algebraically independent differential consequences of~$\mathcal L$
that have, as differential equations, orders not greater than $k$. 
We identify~$L_{(k)}$ with the corresponding system of algebraic equations in~$J^k(x|u)$ and 
associate it with the manifold~$\mathcal L_{(k)}$ determined by this system. 

Here $J^k(x|u)$ is the $k$-th order jet space with the independent variables $x$ and the dependent variables $u$. 
A smooth function defined on a subset of~$J^k(x|u)$ for some $k$, i.e., 
depending on~$x$ and a finite number of derivatives of~$u$, will be called a differential function of~$u$. 
The notation $H[u]$ means that $H$ is a differential function of~$u$. 
See~\cite{Olver1993} for complete definitions.

For the manifold~$\mathcal L_{(k)}$ to actually represent the system~$\mathcal L$ of differential equations, 
the~$\mathcal L$ have to be \emph{locally solvable} in each point of~$\mathcal L_{(k)}$. 
For the application of the Hadamard lemma to differential functions vanishing on the manifold~$\mathcal L_{(k)}$, 
we need the system~$L_{(k)}$ to be, as a system of algebraic equations defined in the jet space~$J^k(x|u)$,
\emph{of maximal rank} in each point of~$\mathcal L_{(k)}$. 
If for any~$k$ the system~$\mathcal L$ satisfies both these conditions then it is called 
\emph{totally nondegenerate}. 
(This definition slightly differs from that given in~\cite{Olver1993}.)

For certain purposes, e.g., for different potential and pseudo-potential frames, 
it is useful to introduce the more general notion of \emph{weight} 
of differential variables instead of the order, which takes into account the structure 
of the system of differential equations under consideration. 
Namely, for each variable of the infinite-order jet space $J^\infty(x|u)$ 
(being the inverse limit of the jet space tower $\{J^k(x|u),k\in\mathbb N\cup\{0\}\}$ 
with respect to the canonical projections $\pi^k\colon J^k(x|u)\to J^{k-1}(x|u)$, $k\in\mathbb N$)
we define its weight~$\varrho$ by the rule: 
\[
\varrho(x_i)=0,\quad \varrho(u^a_\alpha)=\varrho_a+|\alpha|.
\]
The weights $\varrho(u^a)=\varrho_a$ are defined on the basis of the structure of the system~$\mathcal L$. 
(In subsequent sections we will provide concrete examples on how to specify the $\varrho_a$ initially.)
In what follows $u^a_\alpha$ stands for the variable in $J^\infty(x|u)$,
corresponding to the derivative $\p^{|\alpha|}u^a/\p x_1^{\alpha_1}\ldots\p x_n^{\alpha_n}$,
$\alpha=(\alpha_1,\ldots,\alpha_n)$ is an arbitrary multiindex, 
$\alpha_i\in\mathbb{N}\cup\{0\}$, $|\alpha|:=\alpha_1+\cdots+\alpha_n$.
If $\varrho_a=0$ then the weight of~$u^a_\alpha$ obviously coincides with the usual derivative order~$|\alpha|$.
We include in the \emph{weighted jet space} $J^k_\varrho(x|u)$ the variables whose weight is not greater than~$k$.
The infinite-order jet space $J^\infty(x|u)$ is the inverse limit of the weighted jet space tower $\{J^k_\varrho(x|u),\,k\in\mathbb N\cup\{0\}\}$ 
with respect to the canonical projections $\pi^k_\varrho\colon J^k_\varrho(x|u)\to J^{k-1}_\varrho(x|u)$, $k\in\mathbb N$.

The technique of working with weights does not differ from the order technique and so a 
number of analogous notions can be introduced.
Thus, the weight $\varrho(H)$ of any differential function $H[u]$ equals the maximal weight of variables explicitly appearing in~$H$. 
The weight of the equation $H[u]=0$ equals $\varrho(H)$. 
A complete set of independent differential consequences of the system~$\mathcal L$ 
which have weights not greater than $k$ and the associated manifold in $J^k_\varrho(x|u)$ 
are denoted by the symbols $L_{[k]}=L_{[k],\varrho}$ and $\mathcal L_{[k]}=\mathcal L_{[k],\varrho}$, respectively.
The system~$\mathcal L$ is called \emph{totally nondegenerate with respect to the weight~$\varrho$} if for any $k\in\mathbb N$
it is locally solvable in each point of~$\mathcal L_{[k]}$ and 
the algebraic system~$L_{[k]}$ is \emph{of maximal rank} in each point of~$\mathcal L_{[k]}$. 
The Hadamard lemma can be applied, in the conventional way, to differential functions defined in $J^k_\varrho(x|u)$ and vanishing on~$\mathcal L_{[k]}$. 

We will explicitly indicate all places in which the usage of weighted jet spaces is essential. 
In the other places, the terminology involving orders is used although it can be replaced by that based on weights. 

\begin{definition}\label{def.conservation.law}
A {\em conserved vector} of the system~$\mathcal L$ is
an $n$-tuple $F=(F^1[u],\ldots,F^n[u])$ for which the total divergence ${\rm Div}\,F:=D_iF^i$
vanishes for all solutions of~$\mathcal L$, i.e., ${\rm Div}F\bigl|_\mathcal L=0$.
\end{definition}

In Definition~\ref{def.conservation.law} and below
$D_i=D_{x_i}$ denotes the operator of total differentiation with respect to the variable~$x_i$, i.e.
$D_i=\p_{x_i}+u^a_{\alpha+\delta_i}\p_{u^a_\alpha}$, where
$\delta_i$ is the multiindex whose $i$-th entry equals 1 and whose other entries are zero. 
We use the summation convention for repeated indices and consider any function as its zero-order derivative.
The indices $i$ and $j$ run from~1 to~$n$, the index~$a$ runs from 1 to~$m$, and the index~$s$ from 1 to~$p$ unless otherwise stated.
The notation~$V\bigl|_\mathcal L$ means that values of $V$ are considered
only on solutions of the system~$\mathcal L$.

Heuristically, a conservation law of the system~$\mathcal L$ 
is an expression $\mathop{\rm Div}\nolimits F$ vanishing on the solutions of~$\mathcal L$. 
The more rigorous definition of conservation laws given below is based on the factorization of the space 
of conserved vectors with respect to the subspace of trivial conserved vectors. 
Note that there is also a formalized definition of conservation laws of~$\mathcal L$ 
as $(n-1)$-dimensional cohomology classes 
in the so-called horizontal de Rham complex on the infinite prolongation of the system~$\mathcal L$ 
\cite{Bocharov&Co1997,Tsujishita1982,Vinogradov1984}. 
The formalized definition is appropriate for certain theoretical considerations and 
reduces to the usual one after local coordinates are fixed. 

\begin{definition}
A conserved vector $F$ is called {\em trivial} if $F^i=\hat F^i+\check F^i$ 
where $\hat F^i$ and $\check F^i$ are, like $F^i$, differential functions of $u$,
$\hat F^i$ vanishes on the solutions of~$\mathcal L$ and the $n$-tuple $\check F=(\check F^1,\ldots,\check F^n)$
is a null divergence (i.e. its divergence vanishes identically).
\end{definition}

The triviality effected by conserved vectors vanishing on solutions of the system can easily be
eliminated by restricting to the manifold of the system, taking into account all its relevant differential consequences.
A characterization of all null divergences is given by the following theorem (see e.g.~\cite[Theorem~4.24]{Olver1993}).

\begin{theorem}\label{TheoremOnNullDivergences}
The $n$-tuple $F=(F^1,\ldots,F^n)$, $n\ge2$, is a null divergence $(\mathop{\rm Div}\nolimits F\equiv0)$
if and only if there exist differential functions $v^{ij}[u]$ such that $v^{ij}=-v^{ji}$ and $F^i=D_jv^{ij}$.
\end{theorem}

If $n=1$ any null divergence is constant.

\begin{definition}\label{DefinitionOfConsVectorEquivalence}
Two conserved vectors $F$ and $F'$ are called {\em equivalent} if the tuple $F'-F$ is a trivial conserved vector.
\end{definition}

The above definitions of triviality and equivalence of conserved vectors are natural
in view of the usual ``empiric'' definition of conservation laws of a system of differential equations
as divergences of its conserved vectors, i.e., divergence expressions which vanish for all solutions of this system.
For example, equivalent conserved vectors correspond to the same conservation law.
This allows us to formulate the definition of conservation law in a rigorous style (see e.g.~\cite{Zharinov1986}).
Namely, for any system~$\mathcal L$ of differential equations the set~$\CV(\mathcal L)$ of conserved vectors of
its conservation laws is a linear space,
and the subset~$\CV_0(\mathcal L)$ of trivial conserved vectors is a linear subspace in~$\CV(\mathcal L)$.
The factor space~$\CL(\mathcal L)=\CV(\mathcal L)/\CV_0(\mathcal L)$
coincides with the set of equivalence classes of~$\CV(\mathcal L)$ with respect to the equivalence relation adduced in
Definition~\ref{DefinitionOfConsVectorEquivalence}.

\begin{definition}\label{DefinitionOfConsLaws}
The elements of~$\CL(\mathcal L)$ are called {\em (local) conservation laws} of the system~$\mathcal L$,
and the factor space~$\CL(\mathcal L)$ itself is called {\em the space of (local) conservation laws} of~$\mathcal L$.
\end{definition}

This is why we view the determination of the set of conservation laws of~$\mathcal L$ 
as finding~$\CL(\mathcal L)$, which in turn is equivalent to constructing either a basis if
$\dim \CL(\mathcal L)<\infty$ or a system of generators in the infinite dimensional case.
All elements of~$\CV(\mathcal L)$ which belong to the same equivalence class determining a conservation law~$\mathcal F$
are considered as conserved vectors of this conservation law,
and we will additionally identify elements from~$\CL(\mathcal L)$ with their representatives
in~$\CV(\mathcal L)$.
For $F\in\CV(\mathcal L)$ and $\mathcal F\in\CL(\mathcal L)$
the notation~$F\in \mathcal F$ will mean that $F$ is a conserved vector corresponding
to the conservation law~$\mathcal F$.
In contrast to the order $\ord F$ of a conserved vector~$F$ as the maximal order of derivatives explicitly appearing in~$F$,
the {\em order $\ord\mathcal F$ of the conservation law}~$\mathcal F$
is defined as $\min\{\ord F\,|\,F\in\mathcal F\}$.
The notion of weight of a conservation law is introduced in the same way. 
By linear dependence of conservation laws we mean linear dependence as elements of~$\CL(\mathcal L)$.
Therefore, in the framework of the ``representative'' approach
conservation laws of a system~$\mathcal L$ are considered {\em linearly dependent} if
there exists a linear combination of their representatives which is a trivial conserved vector.

Substituting any solution~$u$ of~$\mathcal L$ into any conserved vector~$F$ results in a null divergence depending only on~$x$. 
Then the functions $v^{ij}$ of~$x$, introduced according to Theorem~\ref{TheoremOnNullDivergences} and implicitly parameterized by~$u$,
are called {\em potentials} corresponding to the conserved vector~$F$. 
The equations $D_jv^{ij}=F^i$ determine each potential~$v^{ij}$ up to the negligible summand $\breve v^{ij}$, where 
$\breve v^{ij}=-\breve v^{ji}$ and $D_j\breve v^{ij}=0$. 
Acting on the potentials, the gauge transformation $\tilde v^{ij}=v^{ij}+\breve v^{ij}$ has no influence on 
the corresponding tuple~$F$. 
This gives constant and functional indeterminacies in the potentials if $n=2$ and $n\geqslant3$, respectively. 

\looseness=-1
Suppose that $F$ and $\tilde F$ are equivalent conserved vectors, i.e., 
there exist a null divergence $\check F$ and a tuple $\hat F$ vanishing on the solutions of~$\mathcal L$ such that 
$\tilde F=F+\check F+\hat F$. 
In view of Theorem~\ref{TheoremOnNullDivergences} we can represent $\check F$ in the form $\check F^i=D_j\check v^{ij}$ 
for some differential functions $\check v^{ij}[u]=-\check v^{ji}[u]$.
Then the tuples of potentials $(v^{ij})$ and $(\tilde v^{ij})$ respectively associated with the conserved vectors $F$ and $\tilde F$ 
are connected, up to negligible summands $\breve v^{ij}$, via the transformation $\tilde v^{ij}=v^{ij}+\check v^{ij}[u]$ 
which allows us to assume that these tuples of potentials are equivalent. 
Therefore, we can say that the tuple $(v^{ij})$ (or $(\tilde v^{ij})$) of potentials is associated with the conservation law containing 
the conserved vectors $F$ and~$\tilde F$.

\section{Characteristics of conservation laws}\label{SectionOnCharacteristicsOfConsLaws}

Let the system~$\mathcal L$ be totally nondegenerate.
Then an application of the Hadamard lemma to the definition of conserved vector and integration by parts imply that 
the divergence of any conserved vector of~$\mathcal L$ can always be represented,
up to the equivalence relation of conserved vectors,
as a linear combination of the left hand sides of the independent equations from $\mathcal L$
with coefficients~$\lambda^\mu$ which are functions on a suitable jet space~$J^k(x|u)$:
\begin{equation}\label{EqCharFormOfConsLaw}
\mathop{\rm Div}\nolimits F=\lambda^\mu L^\mu.
\end{equation}
Here the order~$k$ is determined by~$\mathcal L$ and the order of~$F$,
$\mu=\overline{1,l}$. 
More precisely, the following statement is true. 

\begin{proposition}\label{PropositionOnCharRepresentationOfCVs}
For any conserved vector~$F$ of~$\mathcal L$ there exist a tuple $\hat F=(\hat F^1[u],\dots,\hat F^n[u])$ 
vanishing on the solutions of~$\mathcal L$ 
and differential functions $\lambda^\mu[u]$ such that
\[
\mathop{\rm Div}\nolimits F=\lambda^\mu L^\mu+\mathop{\rm Div}\nolimits\hat F.
\]
\end{proposition}

If a tuple  $F=(F^1[u],\dots,F^n[u])$ satisfies equality~\eqref{EqCharFormOfConsLaw} for some differential functions $\lambda^\mu[u]$ 
then it obviously is a conserved vector of~$\mathcal L$.

\begin{definition}\label{DefCharForm}
Formula~\eqref{EqCharFormOfConsLaw} and the $l$-tuple $\lambda=(\lambda^1,\ldots,\lambda^l)$
are called the {\it characteristic form} and the {\it characteristic}
of the conservation law containing the conserved vector~$F$, respectively.
\end{definition}

The characteristic~$\lambda$ is {\em trivial} if it vanishes for all solutions of $\mathcal L$.
Since $\mathcal L$ is nondegenerate, the characteristics~$\lambda$ and~$\tilde\lambda$ satisfy~\eqref{EqCharFormOfConsLaw}
for the same~$F$ and, therefore, are called {\em equivalent}
iff $\lambda-\tilde\lambda$ is a trivial characteristic.
Similarly to conserved vectors, the set~$\Ch(\mathcal L)$ of characteristics
corresponding to conservation laws of the system~$\mathcal L$ is a linear space,
and the subset~$\Ch_0(\mathcal L)$ of trivial characteristics is a linear subspace in~$\Ch(\mathcal L)$.
The factor space~$\Ch_{\rm f}(\mathcal L)=\Ch(\mathcal L)/\Ch_0(\mathcal L)$
coincides with the set of equivalence classes of~$\Ch(\mathcal L)$
with respect to the above characteristic equivalence relation.

We should like to emphasize that the explicit form of characteristics depends on what set of equations is chosen for 
the canonical representation of the system~$\mathcal L$.

The following result~\cite{Olver1993} forms the cornerstone for the methods of studying conservation laws,
which are based on formula~\eqref{EqCharFormOfConsLaw}, including the Noether theorem and
the direct method in the version by Anco and Bluman~\cite{Anco&Bluman2002a,Anco&Bluman2002b}.

\begin{theorem}\label{TheoremIsomorphismChCV}
Let~$\mathcal L$ be a normal, totally nondegenerate system of differential equations.
Then the representation of the conservation laws of~$\mathcal L$ in the characteristic form~\eqref{EqCharFormOfConsLaw}
generates a linear isomorphism between~$\CL(\mathcal L)$ and~$\Ch_{\rm f}(\mathcal L)$.
\end{theorem}

Using properties of total divergences, we can eliminate the conserved vector~$F$ from~\eqref{EqCharFormOfConsLaw}
and obtain a condition for the characteristic~$\lambda$ only.
Namely, a differential function~$f$ is a total divergence, i.e. $f=\mathop{\rm Div} F$
for some $n$-tuple~$F$ of differential functions iff $\Eop(f)=0$.
Here the Euler operator~$\Eop=(\Eop_1,\ldots, \Eop_m)$ is the $m$-tuple of differential operators
\[
\Eop_a=(-D)^\alpha\p_{u^a_\alpha}, \quad a=\overline{1,m},
\]
where
$\alpha=(\alpha_1,\ldots,\alpha_n)$ runs through the multi-index set ($\alpha_i\!\in\!\mathbb{N}\cup\{0\}$),
$(-D)^\alpha=(-D_1)^{\alpha_1}\ldots(-D_m)^{\alpha_m}$.
Therefore, the action of the Euler operator on~\eqref{EqCharFormOfConsLaw} results in the equation
\begin{equation}\label{EqNSCondOnChar}
\Eop(\lambda^\mu L^\mu)={\Fder}_\lambda^*(L)+{\Fder}_L^*(\lambda)=0,
\end{equation}
which is a necessary and sufficient condition on characteristics of conservation laws for the system~$\mathcal L$.
The matrix differential operators~${\Fder}_\lambda^*$ and~${\Fder}_L^*$ are the adjoints of
the Fr\'echet derivatives~${\Fder}_\lambda^{\phantom{*}}$ and~${\Fder}_L^{\phantom{*}}$, i.e.,
\[
{\Fder}_\lambda^*(L)=\left((-D)^\alpha\left( \dfrac{\p\lambda^\mu}{\p u^a_\alpha}L^\mu\right)\right), \qquad
{\Fder}_L^*(\lambda)=\left((-D)^\alpha\left( \dfrac{\p L^\mu}{\p u^a_\alpha}\lambda^\mu\right)\right).
\]
Since ${\Fder}_\lambda^*(L)=0$ automatically on solutions of~$\mathcal L$ then
equation~\eqref{EqNSCondOnChar} implies a necessary condition for $\lambda$ to belong to~$\Ch(\mathcal L)$:
\begin{equation}\label{EqNCondOnChar}
{\Fder}_L^*(\lambda)\bigl|_{\mathcal L}=0.
\end{equation}
Condition~\eqref{EqNCondOnChar} can be considered as adjoint to the criterion
${\Fder}_L^{\phantom{*}}(\eta)\bigl|_{\mathcal L}=0$ for infinitesimal invariance of $\mathcal L$
with respect to an evolutionary vector field with characteristic~$\eta=(\eta^1,\ldots,\eta^m)$.
This is why solutions of~\eqref{EqNCondOnChar} are sometimes called
{\em cosymmetries}~\cite{Sergyeyev2000,Blaszak1998} or
{\em adjoint symmetries}~\cite{Anco&Bluman2002b}.

For the investigation of the connection between characteristics and conserved vectors via formula~\eqref{EqCharFormOfConsLaw}, 
we need a statement on solutions of the equation $D_iF^i=H$, 
where $H=H[u]$ is a given differential function and the $F^i=F^i[u]$ are unknown 
(cf. formula (5.151) and Theorem 5.104 of~\cite{Olver1993}).

\begin{theorem}\label{TheoremOnEqsWithDivergence}
Any solution $F=(F^1,\ldots,F^n)$ of the equation $D_iF^i[u]=H[u]$ 
can be represented in the form $F=\check F+\tilde F$, where 
the $n$-tuple $\check F[u]$ is a null divergence ($D_i\check F^i=0$) and 
the $n$-tuple $\tilde F[u]$ is the particular solution of this equation whose components are given by 
\[
\tilde F^i=\int_0^1\frac{\alpha_i+1}{|\alpha|+1}D^\alpha\bigl(u^a\Eop_a^{\alpha+\delta_i}(H)[\kappa u]\bigr)d\kappa+
\int_0^1x^iH(\kappa x,0,\dots,0)d\kappa.
\]
\end{theorem}
Here $\Eop_a^\alpha$ is the higher-order Euler operator acting on an arbitrary differential function $P[u]$ according to
\[
\Eop_a^\alpha(P)=\sum_{\beta\geqslant\alpha}\frac{\beta!}{\alpha!(\beta-\alpha)!}(-D)^{\beta-\alpha}\frac{\p P}{\p u^a_\beta}.
\] 
Recall also that for any multiindex $\alpha$ with components $\alpha_1,\dots,\alpha_n\in\mathbb{N}\cup\{0\}$, we have 
$\alpha!:=\alpha_1!\cdots\alpha_n!$ and $\delta_i$ was introduced after Definition~\ref{def.conservation.law}.
The condition $\beta\geqslant\alpha$ 
for the multiindices $\alpha=(\alpha_1,\dots,\alpha_n)$ and $\beta=(\beta_1,\dots,\beta_n)$ means that 
$\beta_1\geqslant\alpha_1$, \dots, $\beta_n\geqslant\alpha_n$.

In fact, we need only a consequence of Theorem~\ref{TheoremOnEqsWithDivergence}. 
It is easy to see that if the function~$H$ does not depend on the derivatives of $u^a$ for a fixed value of~$a$ then
the tuple~$\tilde F$ from Theorem~\ref{TheoremOnEqsWithDivergence} possesses the same property with the same value of~$a$.

\begin{corollary}\label{CorollaryOnDependenceOfConservedVectorsOnPartOfVars}
Let $F$ be a conserved vector of a system~$\mathcal L$, satisfying the equality $D_iF^i=H$, where 
the differential function~$H[u]$ does not depend on the derivatives of $u^{a_1}$,\dots, $u^{a_q}$ for fixed values~$a_1$, \dots, $a_q$. 
Then the conserved vector $F$ is equivalent to a conserved vector of~$\mathcal L$ 
which does not depend on the derivatives of $u^{a_1}$,\dots, $u^{a_q}$.
\end{corollary}


\section{A Hadamard lemma for fiber bundles}\label{SectionOnHadamardLemmaForFiberBundles}

In this section we derive certain versions of the well-known Hadamard lemma 
(see e.g., \cite[Proposition~2.10]{Olver1993}) which will be needed in our further investigations.
To this end we will employ the following notations: let $k,\kappa\in\mathbb N$. 
The index~$s$ will run from~1 to $k$, the index~$S$ from~1 to $K$ and the index~$\sigma$ from~1 to $\kappa$. 
Let us also recall that the summation convention for summation over repeated 
indices is in effect.

To begin with we treat a rather elementary special case of the general result below
in order to make the underlying ideas transparent and to single out a case of
practical relevance. In both cases, we will use unified notations. 

Suppose that $B$ and $N$ are manifolds. (Here $N$ can also be a one-element set.) 
Denote the manifold $B\times N\times\mathbb R^\kappa$ by $M$.
Consider the smooth functions $g\colon B\to\mathbb R^k$, $\zeta\colon B\times N\to\mathbb R^\kappa$ and $f\colon B\to\mathbb R$. 
We associate the function~$f$ with the function $\hat f\colon M\to\mathbb R$ 
defined by 
\[
\hat f(y,z',z'')=f(y) \quad \forall(y,z',z'')\in M.
\]

\begin{lemma}\label{LemmaSimplestVersionOfHadamardLemmaForFiberBundles}
Let $g\colon B\to\mathbb R^k$ be a mapping of maximal rank on the submanifold $B_g=\{y\in B\mid g(y)=0\}$. 
The function~$\hat f$ vanishes on the submanifold 
\[M_{g,h}=\{(y,z',z'')\in M\mid g(y)=0,\,h(y,z',z''):=z''-\zeta(y,z')=0\}\] 
if and only if there exists a smooth function $\lambda\colon B\to\mathbb R^k$ such that 
\[
f(y)=\lambda^s(y)g^s(y) \quad \forall y\in B.
\]
\end{lemma}

\begin{proof}
Suppose that the function $\hat f$ vanishes on $M_{g,h}$. 
We fix an arbitrary point $y_0$ from $B_g$ and some point $z'_0$ from~$N$ and put $z''_0=\zeta(y_0,z'_0)$. 
The point $(y_0,z'_0,z''_0)$ of $M$ belongs to $M_{g,h}$ and hence $f(y_0)=\hat f(y_0,z'_0,z''_0)=0$. 
In other words, the function $f$ vanishes on the entire submanifold $B_g$. 
Then the Hadamard lemma implies the desired result.

The converse statement is obvious.
\end{proof}

Lemma~\ref{LemmaSimplestVersionOfHadamardLemmaForFiberBundles} in fact deals with systems of algebraic equations on trivial fiber bundles, 
which are partitioned into two subsets of equations. 
For each appropriate system, the equations of the first subset are pullbacks of equations on the base~$B$ of the fiber bundle under consideration. 
The equations from the second set essentially depends on ``fiber variables'', 
i.e., any nonzero combination of them is not an equation of the first kind.  
Such system can be called \emph{trivially foliated} since the partition is the same for all points of the fiber bundle.
In fact, we can weaken the condition of trivial foliation and demand 
for systems to have at least local representations as pairs of subsystems with the properties described.

Our next aim is to generalize this result to the general fiber bundle setting. To this end we first
introduce some notation (cf., e.g., \cite{Dieudonne1972}).

Consider a smooth fiber bundle $(M,B,\pi,F)$, where $M$ is the total space of the bundle, 
$B$ the base space, $F$ the fiber, and $\pi\colon M\to B$ the projection map. 
We write $(U,\varphi)$ for the local trivializations (or fiber bundle charts) of the bundle~$M$, 
$\smash{\pi^{-1}(U)\stackrel{\varphi}\simeq U\times F}$.
Any point $x\in\pi^{-1}(U)$ corresponds to the pair $(y,z)=\varphi(x)\in B\times F$, i.e., 
$y=\pi(x)=\pr_1(\varphi(x))\in B$ and $z=\pr_2(\varphi(x))\in F$.

Let $H\colon M\to\mathbb R^K$, $g\colon B\to\mathbb R^k$ and $f\colon B\to\mathbb R$ be smooth maps. 
$B_g=\{y\in B\mid g(y)=0\}$ and  $M_H=\{x\in M\mid H(x)=0\}$ denote the set of solutions of the systems 
$g(y)=0$ and $H(x)=0$, respectively. 
We associate the functions~$f$ and~$g$ with their pullbacks $f\circ\pi\colon M\to\mathbb R$ and $g\circ\pi\colon M\to\mathbb R^k$ under $\pi$.

\begin{lemma}\label{LemmaGeneralVersionOfHadamardLemmaForFiberBundles}
Suppose that  $\pi(M_H)=B_g$ and 
$g\colon B\to\mathbb R^k$ has maximal rank on~$B_g$.
Then the function~$f\circ\pi$ vanishes on~$M_H$ 
if and only if there exists a smooth map $\lambda\colon B\to\mathbb R^k$ such that 
\begin{equation}\label{EqHadamardClaim}
f(y)=\lambda^s(y)g^s(y) \quad \forall y\in B.
\end{equation}
\end{lemma}

\begin{proof}
Suppose that the function $f\circ\pi$ vanishes on $M_H$. 
We fix an arbitrary point $y_0$ from $B_g$. 
The condition $\pi(M_H)\supset B_g$ implies that $M_H\cap\pi^{-1}(y_0)\ne\varnothing$. Let $x_0\in M_H\cap\pi^{-1}(y_0)$. 
Then $f(y_0)=f\circ\pi(x_0)=0$.
In other words, the function $f$ vanishes on the entire set $B_g$. 
In view of the Hadamard lemma we obtain equality~\eqref{EqHadamardClaim}.

Conversely, if the function~$f$ admits a representation of the form~\eqref{EqHadamardClaim}, it vanishes on~$B_g$ and, 
therefore, the function~$f\circ\pi$ vanishes on $\pi^{-1}(B_g)\supset M_H$.  
\end{proof}

\begin{definition}\label{DefinitionOfFoliatedSystemOfAlgEqs} 
Let the smooth maps $H\colon M\to\mathbb R^K$ and $g\colon B\to\mathbb R^k$ have maximal rank on $M_H$ and $B_g$, respectively. 
The system $H(x)=0$ is called a \emph{foliated system} over the \emph{base system} $g(y)=0$ if 
$\pi(M_H)=B_g$.
\end{definition}

Definition~\ref{DefinitionOfFoliatedSystemOfAlgEqs} can be reformulated in terms of a connection between the systems $H(x)=0$ and $g(y)=0$. 
This reformulation justifies the name `foliated system'. 

Thus, the condition $\pi(M_H)\subset B_g$ is equivalent to  the pullback of the system $g(y)=0$ being a consequence of the system $H(x)=0$.
Indeed, the condition $\pi(M_H)\subset B_g$ is rewritten as $M_H\subset\pi^{-1}(B_g)$, i.e., the pullback $g\circ\pi$ vanishes on $M_H$. 
By the Hadamard lemma, under the condition of maximal rank of~$H$ on $M_H$ 
there exist functions $\Lambda^{sS}\colon M\to\mathbb R$ such that $g^s\circ\pi(x)=\Lambda^{sS}(x)H^S(x)$. 
This implies that each of the equations $g^s(y)=0$ is a combination of equations of the system $H(x)=0$. 
Conversely, if the system $g\circ\pi(x)=0$ is a consequence of $H(x)=0$, it is obvious that $\pi(M_H)\subset B_g$.

The condition $\pi(M_H)\supset B_g$ means that for any solution~$y_0$ of $g(y)=0$ there exists a solution~$x_0$ of $H(x)=0$ with $\pi(x_0)=y_0$. 
Consider a function $f\colon B\to\mathbb R$ whose pullback $f\circ\pi$ vanishes on~$M_H$. 
Then $f(y_0)=f\circ\pi(x_0)=0$. 
As a result, $f$ vanishes on~$B_g$ and, since the function~$g$ is of maximal rank on $B_g$,
in view of the Hadamard lemma we have $f(y)=\lambda^s(y)g^s(y)$ for some smooth functions $\lambda^s\colon B\to\mathbb R$, 
i.e., the equation $f(y)=0$ is combined from equations of the system $g(y)=0$. 

The above arguments are summarized in the following statement. 

\begin{proposition}\label{PropositionOnFoliatedAlgSystems0}
Suppose that $g\colon B\to\mathbb R^k$ and $H\colon M\to\mathbb R^K$ are smooth mappings 
having maximal rank on the sets $B_g$ and $M_H$, respectively. 
Then the system $H(x)=0$ is foliated over the base system~$g(y)=0$ if and only if 
the pullback $g(\pi(x))=0$ of the system $g(y)=0$ (with respect to the projection~$\pi$) 
is a consequence of the system $H(x)=0$ 
and for any solution~$y_0$ of $g(y)=0$ there exists a solution~$x_0$ of $H(x)=0$ such that $\pi(x_0)=y_0$. 
The foliation also implies that the projection of any combination of equations from the system $H(x)=0$ 
which is the pullback of an equation on~$B$, is a consequence of the system~$g(y)=0$.
\end{proposition}

Let $h\colon M\to\mathbb R^\kappa$ be a smooth map. Then by the {\em vertical rank}
of $h$ in $x\in M$ we mean the rank of the restriction of the tangent map $T_x h$ of $h$
to the vertical subspace of the tangent space $T_xM$ of $M$ at $x$. (This vertical subspace
is just the tangent space of the fiber at $x$.) If $(\varphi,U)$ is any trivialization around
$x$ and $\varphi(x) = (y,z)$, then the vertical rank of $h$ at $x$ is the rank of
$\partial_z (h\circ \varphi^{-1})(y,z)$. After these preparations we may now state:

\begin{theorem}\label{TheremOnLocalStructureOfFoliatedSystems}
Suppose that the system $H(x)=0$ is foliated over the system $g(y)=0$, where $g\colon B\to\mathbb R^k$, $H\colon M\to\mathbb R^K$, 
$k\leqslant\dim B$ and $K\leqslant\dim M$. 
Suppose that $x_0\in M_H$, $y_0=\pi(x_0)$, and $H$ is of constant vertical rank (denoted by~$\kappa$) in a neighborhood of $\pi^{-1}(y_0)\cap M_H$ in~$M$.
Then $K=k+\kappa$ and in a neighborhood~$O_0$ of~$x_0$ in~$M$ the system $H(x)=0$ is equivalent to the united system of $g(\pi(x))=0$ and $h(x)=0$, 
where $h\colon O_0\to\mathbb R^\kappa$ is a smooth function with vertical rank~$\kappa$.
\end{theorem}

\begin{proof}
We choose a fiber bundle chart $(U,\varphi)$ around $y_0$ and set $z_0=\pr_2(\varphi(x_0))$. 
Let $(y^1,\dots,y^n,z^1,\dots,z^m)$ be local coordinates in a neighborhood of $(y_0,z_0)$ in $U\times F$, 
where $n=\dim B$ and $m=\dim F$. Then in what follows we may in fact suppose that $B$ and $F$
are open subsets of $\mathbb R^n$ resp.\ $\mathbb R^m$.
We introduce the notation $y''=(y^1,\dots,y^{K-\kappa})$, $y'=(y^{K-\kappa+1},\dots,y^n)$, $z''=(z^1,\dots,z^\kappa)$ and $z'=(z^{\kappa+1},\dots,z^m)$. 
Up to re-numbering of the $y$- and $z$-variables we can assume that 
$|\p (H\circ\varphi^{-1})/\p (y'',z'')|\ne0$ in the point $(y_0,z_0)$. $H\circ\varphi^{-1}(y_0,z_0)=0$. 
In view of the implicit function theorem there exist neighborhoods $V'$, $V''$, $W'$ and $W''$ of $y'_0$, $y''_0$, $z'_0$ and $z''_0$ 
in the projections of $U\times F$ to the variables $y'$, $y''$, $z'$ and $z''$, respectively, 
and there exist smooth maps $\theta\colon V'\times W'\to V''$ and $\zeta\colon V'\times W'\to W''$ such that 
$H\circ\varphi^{-1}(y,z)=0$ in $\tilde O=V''\times V'\times W''\times W'$ iff $y''=\theta(y',z')$ and $z''=\zeta(y',z')$. 
The derivative $\p\theta/\p z'$ identically vanishes since otherwise $\rank\p H/\p z>\kappa$ for some points in $\tilde O$, 
i.e., in fact $\theta\colon V'\to V''$ and $y''=\theta(y')$.
Note that $y''_0=\theta(y'_0)$ and $z''_0=\zeta(y'_0,z'_0)$.

Since for any solution (in $V''\times V'$) of the system $y''=\theta(y')$ there exists a solution of the system $z''=\zeta(y',z')$ from $W''\times W'$ 
(e.g., $z'=z'_0$ and $z''=\zeta(y',z'_0)$) then 
\[
\pi(M_H\cap O_0)=\{y\in\pi(O_0)\mid y''=\theta(y')\}\subset B_g\cap\pi(O_0), 
\]
where $O_0=\varphi^{-1}(\tilde O)$ and, therefore, $\pi(O_0)=V''\times V'$. 
Consequently, the set of projections of tangent vectors to $M_H$ in the points from $\pi^{-1}(y_0)\cap M_H\cap O_0$ coincides with 
the tangent space to $\pi(M_H\cap O_0)$ in~$y_0$, which has dimension $n-K+\kappa$.

As a result, for any $x\in\pi^{-1}(y_0)\cap M_H$ we can construct a neighborhood $O$ of~$x$ in~$M$ 
such that $\pi(M_H\cap O)\subset B_g\cap\pi(O)$ and 
the set of projections of tangent vectors to $M_H$ in the points from $\pi^{-1}(y_0)\cap M_H\cap O$ 
is an $n-K+\kappa$-dimensional vector space.
It is possible to choose a finite or countable set $\{O_i\}$ of such neighborhoods covering $\pi^{-1}(y_0)\cap M_H$. 
Hence the set of projections of tangent vectors to $M_H$ in the points from $\pi^{-1}(y_0)\cap M_H$ 
is at most a countable union of $n-K+\kappa$-dimensional vector spaces. 
At the same time, it has to coincide with the $n-k$-dimensional tangent space to $B_g$ in~$y_0$ since $\pi(M_H)=B_g$. 
This implies\footnote{If $n-K+\kappa < n-k$ then the Lebesgue measure (in the tangent space $T_{y_0}B_g$) of each of the countably many 
$n-K+\kappa$-dimensional subspaces would be $0$, contradicting the fact that their union is $T_{y_0}B_g$.} 
that $k=K-\kappa$ and, therefore, $\pi(M_H\cap O_0)=B_g\cap\pi(O_0)$, i.e., in view of the Hadamard lemma 
the systems $y''=\theta(y')$ and $g(y)=0$ are equivalent on $\pi(O_0)$. 
Finally, the system $H(x)=0$ is equivalent to the combined system of $g(\pi(x))=0$ and $h(x)=0$ on $O_0$, 
where $h\colon O_0\to\mathbb R^\kappa$ is the smooth function 
defined by $h\circ\varphi^{-1}(y,z)=z''-\zeta(y',z')$ and hence having the vertical rank~$\kappa$.
\end{proof}

\begin{note}
It follows from the proof of Theorem~\ref{TheremOnLocalStructureOfFoliatedSystems} that 
any foliated system 
(under the assumption of constant vertical rank of the associated mapping on the solution submanifold) 
locally has the structure of a trivially foliated system, as treated in Lemma~\ref{LemmaSimplestVersionOfHadamardLemmaForFiberBundles}. 
\end{note}


\section{Foliated systems of differential equations}\label{SectionOnFoliatedSystemsOfDifferentialEquations}

All the potential frames over systems of differential equations investigated in the subsequent sections
are particular cases of the more general notion of foliation of systems of differential equations.

Let~$\bar{\mathcal L}$ be a system~$\bar L(x,u_{(\bar\rho)},v_{(\bar\rho)})=0$ 
of $\bar l$ differential equations $\bar L^1=0$, \ldots, $\bar L^{\bar l}=0$
for $m+p$ unknown functions $u=(u^1,\ldots,u^m)$ and $v=(v^1,\dots,v^p)$
of $n$ independent variables $x=(x_1,\ldots,x_n)$. 
Let~$\mathcal L$ be a system~$L(x,u_{(\rho)})=0$ of $l$ differential equations $L^1=0$, \ldots, $L^l=0$
for only $m$ unknown functions $u$.

For each $k\in\mathbb N\cup\{0\}$ we consider the projection $\varpi_k\colon J^k(x|u,v)\to J^k(x|u)$:  
$\varpi_k(x,u_{(k)},v_{(k)})=(x,u_{(k)})$.
Any differential function $G=G[u]\colon J^k(x|u)\to\mathbb R$ is naturally associated with its pullback 
$\smash{G[u,v]\circ\varpi_k\colon J^k(x|u,v)\to\mathbb R}$ under $\varpi_k$: $G\circ\varpi_k(x,u_{(k)},v_{(k)})=G(x,u_{(k)})$.
It is also possible to consider the projection $\varpi\colon J^\infty(x|u,v)\to J^\infty(x|u)$ whose restriction to $J^k(x|u,v)$ 
coincides with $\varpi_k$ and which induces pullbacks of differential functions of~$u$ of arbitrary (finite) order. 
Usually we will notationally suppress the pullback operation in what follows.
In order to apply, in particular, the usual and extended characteristic forms of conservation laws and the Hadamard lemma, 
we suppose that both the systems~$\mathcal L$ and~$\bar{\mathcal L}$ are totally nondegenerate. 

The definition of foliated systems of differential equations fits well into the general notion of foliation and 
the geometrical interpretation of systems of differential equations as manifolds in a jet space. 

\begin{definition}\label{DefinitionOfFoliatedSystemOfDEs}
The system~$\bar{\mathcal L}$ is called a \emph{foliated system} over the \emph{base system}~$\mathcal L$  
if both the systems~$\mathcal L$ and~$\bar{\mathcal L}$ are totally nondegenerate and 
$\varpi_k(\bar{\mathcal L}_{(k)})=\mathcal L_{(k)}$ for any $k\in\mathbb N$.
\end{definition}

It is natural to denote the relation between~$\bar{\mathcal L}$ and~$\mathcal L$ by $\varpi\bar{\mathcal L}=\mathcal L$. 
Similarly to the algebraic case (cf.\ the previous section), 
Definition~\ref{DefinitionOfFoliatedSystemOfDEs} admits a reformulation in terms of 
a connection between the systems $\bar{\mathcal L}$ and $\mathcal L$, which justifies the name `foliated system'. 
Namely, the system~$\bar{\mathcal L}$ is foliated over the system~$\mathcal L$ if and only 
(the pullback of) each equation of~$\mathcal L$ is a differential consequence of~$\bar{\mathcal L}$ and 
for any local solution $u=u^0(x)$ of~$\mathcal L$ there exist a local solution of the system $\bar L|_{u=u^0}=0$ in~$v$.
The foliation also implies that any differential consequence of~$\bar{\mathcal L}$ which does not involve the functions~$v$ 
is (the pullback of) a differential consequence of~$\mathcal L$.
In terms of solution sets, the strip $u=u^0(x)$, where $u^0(x)$ is a fixed solution of~$\mathcal L$, is 
the solution set of the system $\bar L(x,u^0_{(\bar\rho)},v_{(\bar\rho)})=0$.

\begin{definition}\label{DefinitionOfStronglyFoliatedSystem}
The system~$\bar{\mathcal L}$ is called a \emph{strongly foliated system} over the base system~$\mathcal L$ 
if $\bar{\mathcal L}$ is foliated over~$\mathcal L$ and 
each of the equations minimally representing~$\mathcal L$ can be included in a minimal set of equations forming~$\bar{\mathcal L}$. 
\end{definition}

There exist foliated systems which are not strongly foliated. 
For example, the system~$\bar{\mathcal L}$  formed by the equations $u^2_x=u^1$, $v_x=u^2$ and $v_t=u^1$ is foliated and not strongly foliated 
over the system~$\mathcal L$  consisting of the equations $u^2_x=u^1$ and $u^2_t=u^1_x$. 
Indeed, the equation $u^2_t=u^1_x$ is a differential consequence of~$\bar{\mathcal L}$ and cannot be included 
in the minimal set of equations representing~$\bar{\mathcal L}$. 
The cross differentiation of the two last equations of~$\bar{\mathcal L}$ is the unique way of excluding the derivatives of~$v$ from~$\bar{\mathcal L}$. 
Therefore, any differential consequence of~$\bar{\mathcal L}$ which does not involve the function~$v$ is a differential consequence of~$\mathcal L$. 
This example is directly connected with the main subject of the paper since both the systems are potential systems 
of the $(1+1)$-linear heat equation, cf. systems~\eqref{potsysB0gen} and~\eqref{potsys2B0gen} with the value $A=1$.  

If $\bar{\mathcal L}$ is foliated over~$\mathcal L$, we will assume that the maximally possible number~$\hat l$ of equations of~$\mathcal L$ 
is included in the minimal equation set forming and canonically representing~$\bar{\mathcal L}$. 
Without loss of generality we can additionally assume that these equations are the first $\hat l$ equations in both of these systems. 
Such a representation of~$\bar{\mathcal L}$ and~$\mathcal L$ will be called a \emph{canonical foliation} of~$\bar{\mathcal L}$ over~$\mathcal L$.  
The foliation is strong if and only if $\hat l=l$.

In the previous example we have $\hat l=1$ since the set of equations $u^2_x=u^1$, $v_x=u^2$ and $v_t=u^1$ 
canonically representing~$\bar{\mathcal L}$ includes only one equation ($u^2_x=u^1$) from~$\mathcal L$ 
and cannot include more equations from~$\mathcal L$.  

The pullback of any conserved vector of~$\mathcal L$ under~$\varpi$ obviously is a conserved vector of~$\bar{\mathcal L}$ 
which does not depend on derivatives of~$v$. 
In view of Lemma~\ref{LemmaGeneralVersionOfHadamardLemmaForFiberBundles}, the converse statement is also true. 
Namely, any conserved vector of~$\bar{\mathcal L}$ which does not depend on derivatives of~$v$ is  
the pullback of a conserved vector of~$\mathcal L$ under~$\varpi$.
This justifies the following definition.

\begin{definition}\label{DefinitionOfInducedCLsOfFoliatedSystems}
We say that
a conservation law $\bar{\mathcal F}$ of~$\bar{\mathcal L}$ is a pullback, with respect to~$\varpi$, of a conservation law $\mathcal F$ of~$\mathcal L$
(i.e., $\bar{\mathcal F}=\varpi^*\mathcal F$) 
or, in other words, is \emph{induced} by this conservation law  
if there exists a conserved vector $\bar F\in\bar{\mathcal F}$ which is the pullback of a conserved vector $F\in\mathcal F$. 
\end{definition}

Using Definition~\ref{DefinitionOfInducedCLsOfFoliatedSystems}, we can re-formulate our results on the pullbacks of conserved vectors.

\begin{proposition}\label{PropositionOnInducedCLSOfFoliatedSystems}
A conservation law $\bar{\mathcal F}$ of~$\bar{\mathcal L}$ is induced by a conservation law $\mathcal F$ of~$\mathcal L$ if and only if 
the conservation law $\bar{\mathcal F}$ contains a conserved vector which does not depend on derivatives of~$v$. 
This conserved vector necessarily is the pullback of a conserved vector belonging to~$\mathcal F$. 
\end{proposition}

\begin{definition}\label{DefinitionOfExtendedCharacteristicsOfFoliatedSystems}
Let $\bar{\mathcal L}$ be canonically foliated over~$\mathcal L$. 
A tuple $\lambda=(\lambda^1[u,v],\dots,\lambda^{l+\bar l-\hat l}[u,v])$ is called an \emph{extended characteristic} 
of a conservation law~$\bar{\mathcal F}$ of~$\bar{\mathcal L}$ if some conserved vector $\bar F\in\bar{\mathcal F}$ 
satisfies the condition 
\begin{equation}\label{EqDefOfExtendedChars}
D_i\bar F^i=\sum_{\mu=1}^l\lambda^\mu L^\mu+\sum_{\nu=1}^{\bar l-\hat l}\lambda^{l+\nu}\bar L^{\hat l+\nu}.
\end{equation}
\end{definition}

The definition of usual characteristics involves the minimal set of equations canonically representing 
the system under consideration.
In contrast to this, to define extended characteristics of a canonically foliated system, 
we extend this minimal set by the equations which canonically represent the base system and do not belong to 
the minimal set of equations of the foliated system.

\begin{definition}\label{DefinitionOfInducedExtendedCharacteristicsOfFoliatedSystems}
We say that
a usual or extended characteristic of~$\bar{\mathcal L}$ is \emph{induced} by a characteristic of~$\mathcal L$ 
if the tuple of the characteristic components associated with the pullbacks of the equations of~$\mathcal L$ is 
the pullback of the characteristic of~$\mathcal L$ and the other characteristic components vanish. 
\end{definition}

If the extended characteristic~$\lambda$ is induced by a characteristic of~$\mathcal L$, 
the defining equality~\eqref{EqDefOfExtendedChars} takes the form $D_i\bar F^i=\lambda^\mu[u] L^\mu[u],$ i.e., 
the total divergence of the associated conserved vector~$\bar F$ is a function of only~$x$ and derivatives of~$u$.

\begin{theorem}\label{TheoremOnInducingExtendedCharsOfFoliatedSystems}
Let the system $\bar{\mathcal L}$ be canonically foliated with the base system~${\mathcal L}$.
A conservation law of $\bar{\mathcal L}$ is induced by a conservation law of $\mathcal L$ 
if and only if it has an extended characteristic induced by a characteristic of~$\mathcal L$. 
\end{theorem}

\begin{proof}
Suppose that $\bar{\mathcal F}$ is a conservation law of~$\bar{\mathcal L}$, induced by a conservation law of~$\mathcal L$. 
In view of Proposition~\ref{PropositionOnInducedCLSOfFoliatedSystems}, 
it contains a conserved vector~$\bar F$ which does not depend on derivatives of~$v$. 
The condition $D_i\bar F^i|_{\bar{\mathcal L}}=0$ means that 
the  differential function $D_i\bar F^i$ (of order $r\leqslant\ord(\bar F^1,\dots,\bar F^n)+1$)
vanishes on the manifold $\bar{\mathcal L}_{(r)}$ determined in the jet space $J^r(x|u,v)$ 
by the system~$\bar{\mathcal L}$ and its differential consequences.
Since $\bar{\mathcal L}$ is foliated over~$\mathcal L$ then $\smash{\varpi_r(\bar{\mathcal L}_{(r)})=\mathcal L_{(r)}}$. 
In view of Lemma~\ref{LemmaGeneralVersionOfHadamardLemmaForFiberBundles} 
there exist functions $\breve\lambda^{\breve\mu}$ of only~$x$ and derivatives of~$u$ up to order~$r$ such that 
$D_i\bar F^i=\breve\lambda^{\breve\mu}\breve L^{\breve\mu}$. 
Here the equations $\smash{\breve L^{\breve\mu}=0}$, $\smash{\breve\mu=1,\dots,\breve l}$, 
form a corresponding set of differential consequences of the system~$\mathcal L$ 
which have, as differential equations, order not greater than $r$.
Following the conventional way of deriving the characteristic form of conservation laws~\cite{Olver1993}, 
we integrate by parts on the right-hand side of the last equality and obtain  
$D_i\tilde F^i=\lambda^\mu L^\mu$, where $\tilde F^i$ and $\lambda^\mu$ are functions of~$x$ and derivatives of~$u$. 
The conserved vectors $\bar F$ and $\tilde F$ are equivalent since their difference vanishes on~$\mathcal L$.  
That is why the tuple $(\lambda^1[u],\dots,\lambda^l[u])$ is a characteristic of the system $\mathcal L$, 
associated with the conserved vector~$\tilde F$ which belongs to the conservation law of~$\mathcal L$, 
inducing $\bar{\mathcal F}$.
Therefore, the tuple $(\lambda^1[u],\dots,\lambda^l[u],\lambda^{l+1}=0,\dots,\lambda^{l+\bar l-\hat l}=0)$ 
is an extended characteristic of the foliated system~$\bar{\mathcal L}$, 
associated with the conservation law~$\bar{\mathcal F}$ and 
induced by the characteristic $(\lambda^1[u],\dots,\lambda^l[u])$ of the base system $\mathcal L$.

Conversely, let the tuple $(\lambda^1,\dots,\lambda^{l+\bar l-\hat l})$ be an extended characteristic of 
the foliated system~$\bar{\mathcal L}$ associated with the conservation law~$\bar{\mathcal F}$, 
induced by the characteristic $(\lambda^1,\dots,\lambda^l)$ of the base system $\mathcal L$. 
This means that $\lambda^1=\lambda^1[u]$, \dots, $\lambda^l=\lambda^l[u]$, $\lambda^{l+1}=0$, \dots, $\lambda^{l+\bar l-\hat l}=0$ 
and there exists a conserved vector $\bar F=\bar F[u,v]\in\bar{\mathcal F}$ such that $D_i\bar F^i=\lambda^\mu L^\mu$.
Since the right-hand side $\lambda^\mu L^\mu$ depends only on~$x$ and derivatives of~$u$, 
the equality $D_i\bar F^i=\lambda^\mu L^\mu$ implies in view of Corollary~\ref{CorollaryOnDependenceOfConservedVectorsOnPartOfVars} that 
there exists a conserved vector $\tilde F$ of $\bar{\mathcal L}$, 
which depends only on~$x$ and derivatives of~$u$, 
is equivalent to the conserved vectors $\bar F$ and, therefore, belongs to~$\bar{\mathcal F}$. 
This in turn shows in view of Proposition~\ref{PropositionOnInducedCLSOfFoliatedSystems}
that the conservation law~$\bar{\mathcal F}$ is induced by a conservation law of the base system~$\mathcal L$.
\end{proof}

The proof of Theorem~\ref{TheoremOnInducingExtendedCharsOfFoliatedSystems} also implies the following statement. 

\begin{corollary}
An extended characteristic of~$\bar{\mathcal L}$ is induced by a characteristic of~$\mathcal L$ 
if the tuple of the characteristic components associated with the pullbacks of equations of~$\mathcal L$ does not depend 
on derivatives of~$v$ and the other characteristic components vanish. 
\end{corollary}

In the general case the equality $D_i\tilde F^i=\lambda^\mu L^\mu$ is not 
a characteristic form of the conservation law of~$\bar{\mathcal L}$, containing the conserved vector $F$, 
since some equations canonically representing~$\mathcal L$ may lie outside of the canonical foliation~$\bar{\mathcal L}$.
The strong foliation guaranties the inclusion of all the equations $L^1=0$, \ldots, $L^l=0$ in 
the canonical foliation. 

\begin{corollary}
A conservation law of the canonically strongly foliated system~$\bar{\mathcal L}$ 
is induced by a conservation law of the base system~$\mathcal L$ if and only if 
it has a characteristic induced by a characteristic of~$\mathcal L$. 
\end{corollary}


\section{The two-dimensional case}\label{SectionOn2DimCase}

In this section we first derive our results for the case of two independent variables 
to explain some necessary notions and ideas of the proof clearly.
Moreover, this case is special, in particular, with respect to a possible (constant) indeterminacy
after the introduction of potentials and due to the high effectiveness of the application of potential symmetries.
Only in this case the introduction, according to Theorem~\ref{TheoremOnNullDivergences}, 
of potentials with an arbitrary finite set of conservation laws results in an Abelian covering of the system under consideration, 
and any Abelian covering can be obtain in this way. 

We denote the independent variables by $t$ and $x$.
A conserved vector of the system~$\mathcal L$ in two independent variables~$t$ and~$x$ is
a pair $(F,G)$ of functions depending on $t$, $x$ and a (finite) number of derivatives of~$u$, 
whose total divergence vanishes for all solutions of~$\mathcal L$, i.e.
$(D_tF+D_xG)|_\mathcal L{}=0$. 
Here $D_t$ and $D_x$ are the operators of total differentiation with respect to $t$ and $x$, respectively.
The components $F$ and $G$ are called the {\em conserved density} and the {\em flux} of the conserved vector~$(F,G)$.
Two conserved vectors $(F,G)$ and $(F',G')$ are {\em equivalent} and, therefore, associated with the same conservation law if
there exist functions~$\hat F$, $\hat G$ and~$H$ of~$t$, $x$ and derivatives of~$u$ such that
$\hat F$ and $\hat G$ vanish on~$\mathcal L_{(k)}$ for some~$k$~and
$
F'=F+\hat F+D_xH$, $G'=G+\hat G-D_tH$.

Any conserved vector~$(F,G)$ of~$\mathcal L$ allows one to introduce the new dependent (potential) variable~$v$
by means of the equations
\begin{equation}\label{potsys1}
v_x=F,\qquad v_t=-G.
\end{equation}
To construct several potentials in one step,
we have to use a set of conserved vectors associated with linearly independent conservation laws 
since otherwise the potentials will be dependent in the following sense:
there exists a linear combination of the potentials,
which is, up to a negligible constant summand, a differential function of $u$ only 
(see Proposition~\eqref{PropositionOnPotentialsObtainedWithLinearConservationLaws} below).
In the case of two independent variables we can also introduce the more general notion of
potential dependence~\cite{Popovych&Ivanova2004CLsOfNDCEs}.

Let $v^1$, \ldots, $v^p$ be potentials of the system~$\mathcal L$. 
By $\mathcal L_{\rm p}$ we denote the combined system of~$\mathcal L$ 
and the equations determining the set of potentials~$v^1$, \ldots, $v^p$.

\begin{definition}\label{DefinitionOfPotentialDependence}
The potentials $v^1$, \ldots, $v^p$ are called
{\em dependent on the solution set of the system~$\mathcal L$} (or, for short, {\em dependent})
if there exist $r'\in{\mathbb N}$ and a function~$\Omega$ of the variables $t$, $x$, $u_{(r')}$, $v^1$, \ldots, $v^p$
such that $\Omega_{v^s}\ne0$ for some $s$, $1\leqslant s\leqslant p$, and 
$\Omega(t,x,u_{(r')},v^1,\ldots,v^p)=0$ for any solution $(u,v^1,\ldots,v^p)$ of 
(up to gauge transformations, i.e., up to adding constants to the potentials).
\end{definition}

A proof of local dependence or independence of potentials seems rather hopeless for general classes of differential equations 
since it is closely connected with a precise description of the structure of the associated conservation laws.
Examples of such proofs for particular classes of differential equations (diffusion--convection equations and linear parabolic equations)
were presented in~\mbox{\cite{Popovych&Ivanova2004CLsOfNDCEs,Popovych&Kunzinger&Ivanova2008}}.

\begin{proposition}\label{PropositionOnPotentialsObtainedWithLinearConservationLaws}
If conserved vectors of the system~$\mathcal L$ belong to linearly dependent conservation laws then 
the associated potentials are locally dependent on the solution set of~$\mathcal L$.
\end{proposition}

\begin{proof}
Let $(F^s,G^s)$, $s=1,\dots,p$, be conserved vectors of~$\mathcal L$ such that the corresponding conservation laws are linearly dependent. 
This means that $c_sF^s=\hat F+D_xH$, $c_sG^s=\hat G-D_tH$ for some constants $c_s$ and 
some functions~$\hat F$, $\hat G$ and~$H$ of~$t$, $x$ and derivatives of~$u$, 
where $\hat F$ and $\hat G$ vanish on~$\smash{\mathcal L_{(k)}}$ for some~$k$. 
For each $s$, the potential $v^s$ associated with the conserved vector $(F^s,G^s)$ satisfies the equations $v^s_x=F^s$ and $v^s_t=-G^s$.
Therefore, $c_sv^s_x=D_xH+\hat F$ and $c_sv^s_t=D_tH-\hat G$, i.e., $c_sv^s-H=c=\const$ on the solution set of~$\mathcal L$. 
As a result, we obtain that the potentials $v^s$ are locally dependent with $\Omega=c_sv^s-H$. 
(The constant~$c$ is negligible up to gauge transformations of the potentials.)
\end{proof}

\begin{proposition}\label{PropositionOnPotentialDependence}
Suppose that two tuples of potentials are associated with 
tuples of conserved vectors which are equivalent in the following sense: 
Any conserved vector of each tuple is equivalent to a linear combination of conserved vectors from the other tuple. 
Then either both these tuples of potentials are locally dependent or both are locally independent on the solution set of the system~$\mathcal L$. 
Any potential from each of the tuples is a linear combination of potentials from the other tuple with 
an additional summand which is a differential function of the dependent variables of the initial system.
\end{proposition}

It is natural to call tuples of potentials satisfying the conditions of Proposition~\ref{PropositionOnPotentialDependence} 
{\em equivalent}.
Proposition~\ref{PropositionOnPotentialDependence} implies that, up to the equivalence of tuples of potentials, 
any potential system is associated with a subspace of the space of conservation laws of the initial system and 
does not depend on the choice of a basis in this subspace or of the conserved vectors representing the basis elements. 

In the case of a single equation~$\mathcal L$, pairs of equations of the form~\eqref{potsys1} combine into
the complete potential system if at least one of them is associated with a nonsingular characteristic
(since in this case $\mathcal L$ is a differential consequence of this pair).
As a rule, systems of this kind admit a number of nontrivial symmetries and hence are of great interest.
Note that in the case $l=1$ the characteristic $\lambda=\lambda[u]$ is called singular 
if the differential equation $\lambda[u]=0$ has a solution $u=u(x)$. 
The importance of distinguishing between singular and nonsingular characteristics was emphasized by Bluman~\cite{Bluman1993}.

Suppose that the system~$\mathcal L$ has $p$ linearly independent local conservation laws with conserved vectors $(F^s,G^s)$, $s=1,\dots,p$. 
We introduce the potentials $v^1$,~\dots, $v^p$ associated with this tuple of conserved vectors by the formulas
\begin{equation}\label{EqPotPartIn2DcaseForManyCLs}
v^s_x=F^s[u], \quad v^s_t=-G^s[u],
\end{equation}
assuming additionally that these potentials are locally independent on the solution set of the system~$\mathcal L$.
The corresponding potential system~$\mathcal L_{\rm p}$ is canonically represented by the potential part~\eqref{EqPotPartIn2DcaseForManyCLs} and 
the equations of the system~$\mathcal L$ which are not differential consequences of~\eqref{EqPotPartIn2DcaseForManyCLs} 
and the other equations of~$\mathcal L$, taken together. 
This representation is a canonical foliation of the system~$\mathcal L_{\rm p}$ over the system~$\mathcal L$.
Below the index~$\nu$ runs through the set $\mathcal N$ of the numbers of such equations from~$\mathcal L$. 
The index~$\nu'$ runs through the set $\mathcal N'=\{1,\dots,l\}\backslash\mathcal N$. 
Note that the total number of such equations is equal to or greater than $l-p$ but is not necessarily equal to $l-p$. 

By what was said after Proposition \ref{PropositionOnPotentialDependence}, tuples $v=(v^1,\dots,v^p)$ and $\tilde v=(\tilde v^1,\dots,\tilde v^p)$ 
of potentials
associated with the same $p$-dimensional subspace of~the conservation law space $\CL(\mathcal L)$ of~$\mathcal L$ are equivalent.
In other words, the tuples $v$ and $\tilde v$ of potentials are called equivalent if 
there exist differential functions $\Phi^s[u]$ and constants $c_{s\sigma}$ such that $|c_{s\sigma}|\ne0$ and 
the transformation $\tilde v^s=c_{s\sigma}v^\sigma+\Phi^s[u]$ (the variables~$x$ and derivatives of~$u$ are not transformed) maps
the system~$\mathcal L_{\rm p}$ associated with~$v$ to the system~$\tilde{\mathcal L}_{\rm p}$ associated with~$\tilde v$. 
The tuples $(F^s,G^s,s=1,\dots,p)$ and $(\tilde F^s,\tilde G^s,s=1,\dots,p)$ from the potential parts of these  
systems are connected by the formulas 
$(\tilde F^s-c_{s\sigma}F^\sigma-D_t\Phi^s)\bigl|_{\mathcal L}=0$ and 
$(\tilde G^s-c_{s\sigma}G^\sigma+D_x\Phi^s)\bigl|_{\mathcal L}=0$. 
We will also say that the systems~$\mathcal L_{\rm p}$ and~$\tilde{\mathcal L}_{\rm p}$ are 
\emph{equivalent as potential systems} of the system~$\mathcal L$.

In order to use the characteristic form~\eqref{EqCharFormOfConsLaw} of conservation laws 
we need for the systems~$\mathcal L$ and $\mathcal L_{\rm p}$ to be totally nondegenerate in some sense. 
In the general case, it is difficult to derive the total nondegeneracy of~$\mathcal L_{\rm p}$ in the usual sense~\cite{Olver1993} 
from the corresponding property of~$\mathcal L$. 
That is why we use the following trick based on the special structure 
of the potential part~\eqref{EqPotPartIn2DcaseForManyCLs} of~$\mathcal L_{\rm p}$. 
For any $k\in\mathbb N\cup\{0\}$ we replace the usual jet spaces $J^k(x|u)$ and $J^k(x|u,v)$ 
by the weighted jet space $J^k_\varrho(x|u)$ with a predefined weight~$\varrho$ and 
the weighted jet space $J^k_\varrho(x|u,v)$ in which the weight~$\varrho$ is extended to the derivatives of the potentials~$v$  
according to the rule: 
\[
\varrho(v^s_\alpha)=\max\bigl(0,\varrho(F^s)-1,\varrho(G^s)-1\bigr)+|\alpha|.
\]
Note that this rule is not the only possible choice. There are a number of different ways for this extension. 
The main rule for weighting the potentials is that 
the weights of the left-hand sides of equation~\eqref{EqPotPartIn2DcaseForManyCLs} have to be greater than or
equal to the weights of the corresponding right-hand sides.
Recall that the weight $\varrho(H)$ of any differential function $H$ equals 
the maximal weight of the variables explicitly appearing in~$H$. 
For the extension of the weight $\varrho$ to be canonical (up to permutation of potentials) 
in the class of potential systems equivalent to~$\mathcal L_{\rm p}$, 
we have to choose one of the equivalent tuples of potentials which has the minimal value of $\sum_s\varrho(v^s)$. 
The consideration of the pre-weighted space $J^k_\varrho(x|u)$ is necessary for the investigation of hierarchies of potential systems 
since the system~$\mathcal L$ itself may be a potential system of a system with respect to a part of the unknown functions~$u^a$, 
with the other $u$'s as potentials of the previous level. The first step in this recursive procedure is carried out by assigning
the weight $0$ to all variables $u$ of any initial system $\mathcal L$ in a hierarchy of potential systems.

A complete set~$L_{{\rm p}[k]}$ of independent differential consequences of the system~$\mathcal L_{\rm p}$ 
which have extended weights not greater than $k$ is exhausted by the equations 
\begin{gather*}
\breve L^{\breve\mu}=0, \ \breve\mu=1,\dots,\breve l,\quad  
v^s_{(0,j'+1)}=D_x^{j'}F^s, \quad 
v^s_{(i+1,j)}=-D_t^iD_x^jG^s. 
\end{gather*}
Here the equations $\smash{\breve L^{\breve\mu}=0}$, $\smash{\breve\mu=1,\dots,\breve l}$, 
form a complete set~$L_{[k]}$ of independent differential consequences of the system~$\mathcal L$, which have weights not greater than $k$, and
$v^s_{(i,j)}=\p^{i+j}v^s/\p t^i\p x^j$, $i,j\geqslant0$. 
For each~$s$ the indices $j'$ and $(i,j)$ run through the sets in which $i,j,j'\geqslant0$, $\varrho(v^s)+j'<k$ and $\varrho(v^s)+i+j<k$.

It is obvious that for any $k\in\mathbb N$ the system~$L_{{\rm p}[k]}$ is of maximal rank on the manifold~$\mathcal L_{{\rm p}[k]}$ 
in the weighted jet space $J^k_\varrho(x|u,v)$ if and only if the system~$L_{[k]}$ is of maximal rank on the manifold~$\mathcal L_{[k]}$. 
The local solvability of~$\mathcal L_{\rm p}$ follows from the local solvability of~$\mathcal L$ 
and the compatibility conditions for the potential part and implies the local solvability of~$\mathcal L$ 
since~$\mathcal L$ is a subsystem of~$\mathcal L_{\rm p}$. 

As a result, we have the following statement. 

\begin{lemma}\label{LemmaOnTotalNondegeneracy2Dcase}
The system~$\mathcal L$ is totally nondegenerate with respect to a weight if and only if  
the potential system~$\mathcal L_{\rm p}$ is totally nondegenerate with respect to this weight extended to the derivatives of the potentials.
\end{lemma}

Moreover, $\varpi_{[k]}(\mathcal L_{p[k]})=\mathcal L_{[k]}$ for any $k\in\mathbb N$ since 
$L_{p[k]}$ is a trivially foliated system of algebraic equations with the base system $L_{[k]}$.
Therefore, two-dimensional potential systems form a particular case of foliated systems of differential equations and 
all statements of Section~\ref{SectionOnFoliatedSystemsOfDifferentialEquations} are true for conservation laws of such systems. 
(Only in the proof of Theorem~\ref{TheoremOnInducingExtendedCharsOfFoliatedSystems} 
the orders and usual jet spaces have to be replaced by the weights of the same objects and weighted jet spaces, respectively.)
At the same time, due to their special structure stronger statements on the connection between 
conservation laws induced by conservation laws of the corresponding initial systems and 
the locality of the associated characteristics can be proved.

\begin{lemma}\label{LemmaOnCLwithLocalCV2Dcase}
If a characteristic of a two-dimensional potential system depends only on local variables (i.e., independent and non-potential dependent ones) 
then the associated conservation law of the potential system has a conserved vector which also does not depend on potentials. 
\end{lemma}

\begin{proof}
Suppose that the potential system~$\mathcal L_{\rm p}$ possesses a characteristic 
\[
(\alpha^s,\beta^s,\gamma^\nu,\,s=1,\dots,p,\,\nu\in\mathcal N)
\] 
which does not depend on the potentials $v^1$,~\dots, $v^p$. 
(Due to system~\eqref{EqPotPartIn2DcaseForManyCLs} the dependence of the characteristic on nonzero derivatives of the potentials 
can be neglected up to the equivalence relation of characteristics.) 
The components $\alpha^s$, $\beta^s$ and $\gamma^\nu$ are functions of~$x$ and derivatives of~$u$ 
and correspond to the equations $v^s_t=-G^s$, $v^s_x=F^s$ and $L^\nu=0$, respectively. 
Therefore, there exists a conserved vector $(F,G)$ of the potential system~$\mathcal L_{\rm p}$ such that 
\begin{equation}\label{EqCharFormOf2DPotCLa}
D_tF+D_xG=\alpha^s(v^s_t+G^s)+\beta^s(v^s_x-F^s)+\gamma^\nu L^\nu=:V.
\end{equation}
Since the differential function $V$ of~$t$, $x$ and derivatives of~$u$ and~$v$ is a total divergence then 
the value of the extended Euler operator~$\Eop=(\Eop_{u^1},\ldots, \Eop_{u^m},\Eop_{v^1},\ldots, \Eop_{v^p})$ on~$V$ 
is the zero $m+p$-tuple. In particular, 
\[
-\Eop_{v^s}V=D_t\alpha^s+D_x\beta^s=0, 
\]
i.e., the tuple $(\alpha^s[u],\beta^s[u])$ is a null divergence.
In view of Theorem~\ref{TheoremOnNullDivergences} on null divergences, for each~$s$ 
there exists a differential function $\Phi^s[u]$ such that $\alpha^s=D_x\Phi^s$ and $\beta^s=-D_t\Phi^s$.
We set 
\[
\hat F=F+\Phi^s(v^s_x-F^s), \quad \hat G=G-\Phi^s(v^s_t+G^s).
\]
Then equation~\eqref{EqCharFormOf2DPotCLa} can be re-written as
\[
D_t\hat F+D_x\hat G=\Phi^sD_t(v^s_x-F^s)-\Phi^sD_x(v^s_t+G^s)+\gamma^\nu L^\nu=-\Phi^s(D_tF^s+D_xG^s)+\gamma^\nu L^\nu,
\]
and the conserved vector $(\hat F,\hat G)$ is equivalent to the initial conserved vector $(F,G)$.  
The right-hand side of the last equality is a differential function of~$u$ and
vanishes on the manifold~$\mathcal L_{[k]}$ of the jet space~$J^k_\varrho(x|u)$, where 
$k$ is the highest weight of the variables in this expression. 
Using the Hadamard lemma and ``integration by parts'' as in deriving the general characteristic form of 
conservation laws, we obtain that 
\begin{equation}\label{EqCharFormOf2DPotCLb}
D_t\check F+D_x\check G=\check\gamma^\mu L^\mu
\end{equation}
for some differential functions $\check\gamma^\mu[u]$, 
where the conserved vector $(\check F,\check G)$ is equivalent to $(\hat F,\hat G)$ and, therefore, to $(F,G)$ since 
it differs from $(\hat F,\hat G)$ on a tuple vanishing on the solution set of~$\mathcal L$. 
Since the right-hand side $\check\gamma^\mu L^\mu$ depends only on~$x$ and derivatives of~$u$, 
equality~\eqref{EqCharFormOf2DPotCLb} implies in view of Corollary~\ref{CorollaryOnDependenceOfConservedVectorsOnPartOfVars} that 
there exists a conserved vector $(\tilde F,\tilde G)$ of $\mathcal L_{\rm p}$, 
which depends only on~$x$ and derivatives of~$u$ and 
is equivalent to the conserved vectors $(\check F,\check G)$ and, therefore, $(F,G)$.
\end{proof}

\begin{lemma}\label{LemmaOnCLwithLocalChar2Dcase}
If an extended characteristic of a two-dimensional potential system is induced by a characteristic of the corresponding initial system
then the associated conservation law of the potential system has a characteristic which does not depend on potentials. 
\end{lemma}

\begin{proof}
Suppose that the potential system~$\mathcal L_{\rm p}$ possesses an extended characteristic 
induced by a characteristic~$\lambda$ of the initial system~$\mathcal L$, 
i.e., there exists a conserved vector $(F,G)$ of~$\mathcal L_{\rm p}$ such that $D_tF+D_xG=\lambda^\mu[u]L^\mu[u]$.  
In the general case this equality is not a characteristic form 
of the conservation law of~$\mathcal L_{\rm p}$ containing the conserved vector $(F,G)$, 
since some equations of~$\mathcal L$ can fall out of the minimal set of equations 
forming the potential system~$\mathcal L_{\rm p}$.
The indices of such equations form the set~$\mathcal N'=\{\nu'\}$.
If $\mathcal N'=\varnothing$, we at once have a characteristic form. 

Let $\mathcal N'\ne\varnothing$. We represent each $L^{\nu'}$ as a differential consequence of~$\mathcal L_{\rm p}$.
In view of Lemma~\ref{LemmaSimplestVersionOfHadamardLemmaForFiberBundles}, 
this representation has the form 
\[
L^{\nu'}=A^{\nu'\nu}L^\nu+B^{\nu's}(D_tF^s+D_xG^s),
\]
where $A^{\nu'\nu}$ and $B^{\nu's}$ are polynomials of the total differentiation operators $D_t$ and $D_x$ with 
coefficients depending on $t$, $x$ and derivatives of~$u$.
Note that $D_tF^s+D_xG^s=D_x(v^s_t+G^s)-D_t(v^s_x-F^s)$. 
Therefore, 
\[
D_tF+D_xG=\lambda^\nu L^\nu+\lambda^{\nu'}A^{\nu'\nu}L^\nu
+\lambda^{\nu'}B^{\nu's}D_x(v^s+G^s)-\lambda^{\nu'}B^{\nu's}D_t(v^s_x-F^s).
\]
Integrating by parts on the right-hand side leads to the equality
\[
D_t\tilde F+D_x\tilde G=\alpha^s(v^s_t+G^s)+\beta^s(v^s_x-F^s)+\gamma^\nu L^\nu,
\]
where $\alpha^s=-D_xB^{s\nu'*}\lambda^{\nu'}$, $\beta^s=D_tB^{s\nu'*}\lambda^{\nu'}$ and 
$\gamma^\nu=\lambda^\nu+A^{\nu\nu'*}\lambda^{\nu'}$ are functions of~$x$ and derivatives of~$u$, and 
$A^{\nu\nu'*}$ and $B^{s\nu'*}$ denote the formally adjoint operators to $A^{\nu'\nu}$ and $B^{\nu's}$, respectively. 
The conserved vectors $(F,G)$ and $(\tilde F,\tilde G)$ are equivalent since their difference vanishes on $\mathcal L_{\rm p}$.

Finally, we construct the characteristic $(\alpha^s,\beta^s,\gamma^\nu,\,s=1,\dots,p,\,\nu\in\mathcal N)$ 
of the conservation law with the conserved vector $(F,G)$, which depends only on~$x$ and derivatives of~$u$.
\end{proof}

Proposition~\ref{PropositionOnInducedCLSOfFoliatedSystems}, Theorem~\ref{TheoremOnInducingExtendedCharsOfFoliatedSystems} and 
Lemmas~\ref{LemmaOnCLwithLocalCV2Dcase} and~\ref{LemmaOnCLwithLocalChar2Dcase} can now be combined into the following result:

\begin{theorem}\label{TheoremOnCLsOf2DPotSystems}
The following statements on a conservation law of a two-dimensional potential system are equivalent:

1) the conservation law is induced by a conservation law of the corresponding initial system;

2) it contains a conserved vector which does not depend on potentials;

3) some of its extended characteristics are induced by characteristics of the initial system;

4) it possesses a characteristic not depending on potentials.
\end{theorem}

\begin{note}
If~$\mathcal L_{\rm p}$ and~$\tilde{\mathcal L}_{\rm p}$ are equivalent as potential systems of the system~$\mathcal L$ 
then the corresponding equivalence transformation maps any conservation law of~$\mathcal L_{\rm p}$, 
possessing the locality properties 1--4 of Theorem~\ref{TheoremOnCLsOf2DPotSystems}, to a conservation law of~$\tilde{\mathcal L}_{\rm p}$ 
with the same properties. 
In other words, the locality properties of conservation laws are stable with respect to the equivalence of potential systems.
\end{note}

\begin{note}
Although the general version of the Hadamard lemma for fiber bundles (Lemma~\ref{LemmaGeneralVersionOfHadamardLemmaForFiberBundles}) 
is used in the proof of Theorem~\ref{TheoremOnInducingExtendedCharsOfFoliatedSystems} involved in deriving Theorem~\ref{TheoremOnCLsOf2DPotSystems},
in fact the simplest version of this lemma (Lemma~\ref{LemmaSimplestVersionOfHadamardLemmaForFiberBundles}) is sufficient, 
due to the special foliation structure of two-dimensional potential systems, to directly prove Theorem~\ref{TheoremOnCLsOf2DPotSystems}. 
The same observation is true for Abelian coverings and standard potential systems without gauges in the multidimensional case. 
\end{note}

Consider a \emph{tower $\{\mathcal L_{\rm p}^k, k=0,1,\dots,N\}$ of potential systems} over the system $\mathcal L=\mathcal L_{\rm p}^0$.
(Here $N\le \infty$.) 
This means that for any $k\in\{1,\dots,N\}$ the system $\smash{\mathcal L_{\rm p}^k}$ is a potential system of $\mathcal L_{\rm p}^{k-1}$.
The system $\mathcal L_{\rm p}^k$ will be called a \emph{$k$-th level potential system} associated with~$\mathcal L$.
We will say that the potential system $\mathcal L_{\rm p}^k$ is  \emph{strictly of $k$-th level} 
if it cannot be included as a potential system of a lower level in another tower of potential systems over~$\mathcal L$.
For any $k,k'\in\{1,\dots,N\}$, where $k'<k$, the system $\smash{\mathcal L_{\rm p}^k}$ is foliated over the system $\mathcal L_{\rm p}^{k'}$.

A conservation law of a potential system of $k$-th level is a $k$-th level potential conservation law of~$\mathcal L$. 
A conservation law of a potential system which is strictly of $k$-th level and is not induced by a conservation law of lower level, 
is called a \emph{potential conservation law which is strictly of $k$-th level}. 
A potential in a tower of potential systems is \emph{strictly of $k$-th level} 
if it is introduced with a conservation law which is strictly of $(k-1)$-st level.
By linearly combining potentials and lowering their levels as far as possible, 
any finite tower of potential systems over~$\mathcal L$ can be transformed to a tower in which for any $k$ 
the dependent variables of $\mathcal L_{\rm p}^k$, complementary to the dependent variables of $\mathcal L_{\rm p}^{k-1}$, 
are potentials which are strictly of $(k-1)$-st level. 
Another approach to ordering towers of potential systems is to consider only one-dimensional extensions of the spaces of 
dependent variables for each step between levels (see, e.g.,~\cite{Marvan2004}).

An iterative application of Theorem~\ref{TheoremOnCLsOf2DPotSystems} to towers of potentials systems implies two statements 
on potential conservation laws (in terms of a fixed tower and in terms of levels, respectively).

\begin{corollary}\label{CorollaryOnCLsOfTower2DPotSystems}
Let $\{\mathcal L_{\rm p}^k, k=0,1,\dots,N\}$ be a tower of potential systems over the system $\mathcal L$ 
with two independent variables. 
For any $k\in\{1,\dots,N\}$ the following statements on a conservation law of $\mathcal L_{\rm p}^k$ are equivalent:

1) the conservation law is induced by a conservation law of $\mathcal L_{\rm p}^{k'}$ for some $k'<k$;

2) it contains a conserved vector depending only on variables appearing in $\mathcal L_{\rm p}^{k'}$ and derivatives involving them;

3) some of its extended characteristics are induced by characteristics of $\mathcal L_{\rm p}^{k'}$;

4) it possesses a characteristic which does not depend on potentials complementary to the dependent variables of $\mathcal L_{\rm p}^{k'}$.
\end{corollary}

\begin{corollary}\label{CorollaryOn2DPotCLsOfArbitraryLevel}
The following statements on a $k$-th level potential conservation law of a two-dimensional system are equivalent:

1) the conservation law is induced by a conservation law of a lower level;

2) it contains a conserved vector which does not depend on potentials whose strict levels are greater than $k-1$;

3) some of its extended characteristics are induced by characteristics of a potential system of a lower level;

4) it possesses a characteristic not depending on potentials with strict levels greater than $k-1$.
\end{corollary}


\section{Abelian coverings}\label{SectionOnAbelianCoverings}

There are two ways to directly generalize the above results for the two-dimensional case to the multi-dimensional case. 
One of them deals with so-called Abelian coverings~\cite{Marvan2004} 
and the other is based on the introduction of potentials according to Theorem~\ref{TheoremOnNullDivergences}. 
In this section we consider Abelian coverings (in the local approach, 
cf. the remark preceding Definition~\ref{DefinitionOfPotentialDependenceForAbelianCoverings} below).

Suppose that the system~$\mathcal L$ admits $p$ potentials $v^1$,~\dots, $v^p$ defined by the equations 
\begin{equation}\label{EqPotPartInMultiDimCaseForAbelianCoverings}
v^s_i=G^{si}[u], 
\end{equation}
where the differential functions $G^{si}=G^{si}[u]$ satisfy the compatibility conditions $D_jG^{si}=D_iG^{sj}$ on 
the solution set of the system~$\mathcal L$. 
The corresponding potential system~$\mathcal L_{\rm p}$ is canonically represented by 
the potential part~\eqref{EqPotPartInMultiDimCaseForAbelianCoverings} and 
the equations of the system~$\mathcal L$, which are not  differential consequences of~\eqref{EqPotPartInMultiDimCaseForAbelianCoverings} 
and other equations of~$\mathcal L$, taken simultaneously. 
Similarly to Section~\ref{SectionOn2DimCase}, below the index~$\nu$ runs through the set $\mathcal N$ of the numbers of such equations 
and the index~$\nu'$ runs through the complimentary set $\mathcal N'=\{1,\dots,l\}\backslash\mathcal N$. 
The representation described gives a canonical foliation of the system~$\mathcal L_{\rm p}$ over the system~$\mathcal L$. 

The system~$\mathcal L_{\rm p}$ defines a (first level) Abelian covering of the system~$\mathcal L$ since the right-hand sides $G^{si}$ 
of~\eqref{EqPotPartInMultiDimCaseForAbelianCoverings} do not depend on the potentials $v^1$,~\dots, $v^p$.
Each of the compatibility conditions $(D_jG^{si}-D_iG^{sj})\bigl|_{\mathcal L}=0$ 
can be interpreted as a conservation law of~$\mathcal L$ with a conserved vector which 
has only two nonzero components, namely, the $i$-th component equal to~$G^{sj}$ and the $j$-th component equal to~$-G^{si}$. 
Therefore, defining a potential in the framework of Abelian coverings involves $\frac12n(n-1)$ conserved vectors of a special form. 

Similarly to the two-dimensional case, 
two tuples $v=(v^1,\dots,v^p)$ and $\tilde v=(\tilde v^1,\dots,\tilde v^p)$ of potentials of Abelian coverings 
of the same multi-dimensional system~$\mathcal L$ are \emph{equivalent} 
if there exist differential functions $\Phi^s[u]$ and constants $c_{s\sigma}$ such that $|c_{s\sigma}|\ne0$ and 
the transformation $\tilde v^s=c_{s\sigma}v^\sigma+\Phi^s[u]$ (the variables~$x$ and derivatives of~$u$ are not transformed) maps
the system~$\mathcal L_{\rm p}$ associated with~$v$ to the system~$\tilde{\mathcal L}_{\rm p}$ associated with~$\tilde v$. 
The function tuples $(G^{si}[u])$ and $(\tilde G^{si}[u])$ from the potential parts of these systems 
are connected by the formula $(\tilde G^{si}-c_{s\sigma}G^{\sigma i}-D_i\Phi^s)\bigl|_{\mathcal L}=0$. 
In fact, in the local-coordinate approach an Abelian covering of~$\mathcal L$ 
is an equivalence class of tuples of potentials which are considered along with 
the corresponding equations of the form~\eqref{EqPotPartInMultiDimCaseForAbelianCoverings} 
and prolongations of the total differentiation operators to the potentials, coinciding on the solution set of~$\mathcal L$.
The equivalence of potential $p$-tuples agrees with 
the equivalence of the associated $p$-element sets of $\frac12n(n-1)$-tuples of conserved vectors, involving linear combinations. 

Definition~\ref{DefinitionOfPotentialDependence} is easily generalized to Abelian coverings of arbitrary dimensions.

\begin{definition}\label{DefinitionOfPotentialDependenceForAbelianCoverings}
The potentials $v^1$, \ldots, $v^p$ are called
{\em locally dependent on the solution set of the system~$\mathcal L$} (or, briefly speaking, {\em dependent})
if there exist $r'\in{\mathbb N}$ and a function~$\Omega$ of the variables $x$, $u_{(r')}$, $v^1$, \ldots, $v^p$
such that $\Omega_{v^s}\ne0$ for some~$s$ and 
$\Omega(x,u_{(r')},v^1,\ldots,v^p)=0$ for any solution $(u,v^1,\ldots,v^p)$ of the system~$\mathcal L_{\rm p}$, 
(up to gauge transformations, i.e., adding constants to potentials).
\end{definition}

If a linear combination of the tuples $(G^{s1},\dots,G^{sn})$, $s=1,\dots,p$, is a total gradient, i.e., 
$c_sG^{si}=D_iH$ for certain constants $c_s$ and a differential function~$H[u]$, then 
the potentials $v^1$,~\dots, $v^p$ are dependent since $c_sv^s=H[u]+c_0$ for some negligible constant~$c_0$.

Employing the characteristic form~\eqref{EqCharFormOfConsLaw} of conservation laws  
requires the assumption that the systems~$\mathcal L$ and $\mathcal L_{\rm p}$ are totally nondegenerate. 
We again use the trick of introducing weighted jet spaces and extending the weight to potentials.
The procedure is analogous to that in the two-dimensional case. 
Thus, the rule for extending the weight to the derivatives of the potentials $v^1$,~\dots, $v^p$ is  
\[
\varrho(v^s_\alpha)=\max\bigl(0,\varrho(G^{s1})-1,\dots,\varrho(G^{sn})-1\bigr)+|\alpha|.
\]

\begin{lemma}\label{LemmaOnTotalNondegeneracyOfAbelianCoverigs}
The system~$\mathcal L$ is totally nondegenerate with respect to a weight if and only if 
the system~$\mathcal L_{\rm p}$ is totally nondegenerate with respect to this weight extended to the derivatives of the potentials.
\end{lemma}

\begin{proof}
A complete set~$L_{{\rm p}[k]}$ of independent differential consequences of the system~$\mathcal L_{\rm p}$ 
which have extended weights not greater than $k$ is exhausted by the equations 
\begin{gather*}
\breve L^{\breve\mu}=0, \ \breve\mu=1,\dots,\breve l,\quad
v^s_\alpha=D_i^{\alpha_i-1}D_{i+1}^{\alpha_{i+1}}\dots D_n^{\alpha_n}G^{si}.
\end{gather*}
Here the equations $\smash{\breve L^{\breve\mu}=0}$, $\smash{\breve\mu=1,\dots,\breve l}$, 
form a complete set~$L_{[k]}$ of independent differential consequences of the system~$\mathcal L$, 
which have weights not greater than~$k$.
$v^s_\alpha=\p^{|\alpha|}v^s/\p x_1^{\alpha_1}\dots\p x_n^{\alpha_n}$. 
For each~$i$ and $s$ the multiindex $\alpha=(\alpha_1,\dots,\alpha_n)$ runs through the multiindex set in which 
$\alpha_1=\dots=\alpha_{i-1}=0$, $\alpha_i>0$, $\varrho(v^s)+|\alpha|\leqslant k$.

It is obvious that for any $k\in\mathbb N$ the system~$L_{{\rm p}[k]}$ is of maximal rank on the manifold~$\mathcal L_{{\rm p}[k]}$ 
in the weighted jet space $J^k_\varrho(x|u,v)$ if and only if the system~$L_{[k]}$ is of maximal rank on the manifold~$\mathcal L_{[k]}$. 
The local solvability of~$\mathcal L_{\rm p}$ follows from the local solvability of~$\mathcal L$ 
and the compatibility conditions for the potential part and implies the local solvability of~$\mathcal L$ 
since~$\mathcal L$ is a subsystem of~$\mathcal L_{\rm p}$. 
\end{proof}

Since any potential system representing an Abelian covering is foliated over the corresponding initial system, 
all statements of Section~\ref{SectionOnFoliatedSystemsOfDifferentialEquations} are applicable to its conservation laws 
(after the necessary modifications in the proof of Theorem~\ref{TheoremOnInducingExtendedCharsOfFoliatedSystems}, 
connected with the introduction of weighted jet spaces). 
Stronger statements on the connection between potential-free characteristics and conservation laws 
by induced conservation laws of the corresponding initial system can be proved owing to a special structure of the foliation. 

\begin{lemma}\label{LemmaOnCLwithLocalCVMultiDimCaseForAbelianCoverings}
If a characteristic of the potential system~$\mathcal L_{\rm p}$ depends only on ``local'' variables 
(i.e., it is a function only of $x$ and derivatives of~$u$) 
then the associated conservation law of~$\mathcal L_{\rm p}$ has a conserved vector which also does not depend on potentials. 
\end{lemma}

\begin{proof}
Let the potential system~$\mathcal L_{\rm p}$ possess a characteristic 
\[
(\alpha^{si},\gamma^\nu,\,s=1,\dots,p,\,i=1,\dots,n,\,\nu\in\mathcal N)
\] 
which does not depend on the potentials $v^1$,~\dots, $v^p$. 
(By the defining equation~\eqref{EqPotPartInMultiDimCaseForAbelianCoverings} for the potentials, 
the dependence of the characteristic on nonzero derivatives of the potentials 
can be neglected up to the equivalence relation of characteristics.) 
In the above expresseion, $\alpha^{si}$ and $\gamma^\nu$ are differential functions of~$u$ 
corresponding to $v^s_i=G^{si}$ and $L^\nu=0$, respectively. 
From this, we obtain a conserved vector $(F^1,\dots,F^n)$ of $\mathcal L_{\rm p}$ with 
\begin{equation}\label{EqCharFormOfCLsForAbelianCovering}
D_iF^i=\alpha^{si}(v^s_i-G^{si})+\gamma^\nu L^\nu=:V.
\end{equation}
As the differential function $V$ of~$x$ and derivatives of~$u$ and~$v$ is a total divergence,
an application of the extended Euler operator~$\Eop=(\Eop_{u^1},\ldots, \Eop_{u^m},\Eop_{v^1},\ldots, \Eop_{v^p})$ on~$V$ 
gives the zero $m+p$-tuple. We conclude that 
\[
-\Eop_{v^s}V=D_i\alpha^{si}=0, \quad s=1,\dots,p, 
\]
so that $(\alpha^{s1}[u],\dots,\alpha^{sn}[u])$ is a null divergence.
Thus by Theorem~\ref{TheoremOnNullDivergences} 
there exist differential functions $\Phi^{sij}[u]$ such that $\alpha^{si}=D_j\Phi^{sij}$ and $\Phi^{sij}=-\Phi^{sji}$.
Setting 
\[
\hat F^i=F^i+\Phi^{sij}(v^s_j-G^{sj}),
\]
the tuple $\hat F=(\hat F^1,\dots,\hat F^n)$ is a conserved vector equivalent to the initial conserved vector~$F$.
In terms of~$\hat F$ equation~\eqref{EqCharFormOfCLsForAbelianCovering} can be re-written as
\begin{gather*}
D_i\hat F^i=\alpha^{si}(v^s_i-G^{si})+\gamma^\nu L^\nu+(D_i\Phi^{sij})(v^s_j-G^{sj})+\Phi^{sij}(v^s_{ij}-D_iG^{sj})
\\\phantom{D_i\hat F^i}
=\sum_{i<j}\Phi^{sij}(D_jG^{si}-D_iG^{sj})+\gamma^\nu L^\nu.
\end{gather*}
The right-hand side of the last equality vanishes on the solution set of~$\mathcal L$.
The standard way of deriving the characteristic form of conservation laws implies that 
\begin{equation}\label{EqCharFormOfPotCLbForAbelianCovering}
D_i\check F^i=\check\gamma^\mu L^\mu
\end{equation}
for some differential functions $\check\gamma^\mu[u]$ and some conserved vector $\check F$ equivalent to $\hat F$ and, therefore, $F$. 
(The conserved vector $\check F$ differs from $\hat F$ by a tuple vanishing on the solution set of~$\mathcal L$.) 
As $\check\gamma^\mu L^\mu$ depends only on~$x$ and derivatives of~$u$, 
by~\eqref{EqCharFormOf2DPotCLb} and Corollary~\ref{CorollaryOnDependenceOfConservedVectorsOnPartOfVars} 
we obtain that there exist a conserved vector $\tilde F$ of $\mathcal L_{\rm p}$
which depends only on~$x$ and derivatives of~$u$ and
is equivalent to the conserved vector $\check F$ and, consequently, to $F$.
\end{proof}

\begin{lemma}\label{LemmaOnCLwithLocalCharMultiDimCaseForAbelianCoverings}
If an extended characteristic of a potential system~$\mathcal L_{\rm p}$ is induced by a characteristic of the initial system~$\mathcal L$
then the associated conservation law of~$\mathcal L_{\rm p}$ has a characteristic which does not depend on potentials. 
\end{lemma}

\begin{proof}
Let the system~$\mathcal L_{\rm p}$ define an Abelian covering of the system~$\mathcal L$ 
and suppose that $\mathcal L_{\rm p}$ possesses an extended characteristic induced by a characteristic~$\lambda$ of~$\mathcal L$. 
Equivalently, there exists a conserved vector $F$ of~$\mathcal L_{\rm p}$ with $D_iF^i=\lambda^\mu[u]L^\mu[u]$.  
In general, this equation need not be a characteristic form 
of the conservation law of~$\mathcal L_{\rm p}$, containing the conserved vector $F$, 
since some equations of~$\mathcal L$ may fail to be contained in the minimal set of equations 
forming the potential system~$\mathcal L_{\rm p}$.
We collect the indices of such equations in the set~$\mathcal N'=\{\nu'\}$ and suppose that
$\mathcal N'\ne\varnothing$ (as otherwise we already have a characteristic form). 

By Lemma~\ref{LemmaSimplestVersionOfHadamardLemmaForFiberBundles}, 
the representation of any $L^{\nu'}$ as a differential consequence of~$\mathcal L_{\rm p}$
is of the form 
\[
L^{\nu'}=A^{\nu'\nu}L^\nu+\sum_{i<j}B^{\nu'sij}(D_iG^{sj}-D_jG^{si}),
\]
where $A^{\nu'\nu}$ and $B^{\nu'sij}$ are polynomials of the total differentiation operators $D_i$ with 
smooth coefficients depending on~$x$ and derivatives of~$u$, $B^{\nu'sij}=-B^{\nu'sji}$.
Since $D_iG^{sj}-D_jG^{si}=D_j(v^s_i-G^{si})-D_i(v^s_j-G^{sj})$ we get
\[
D_iF^i=\lambda^\nu L^\nu+\lambda^{\nu'}A^{\nu'\nu}L^\nu+\lambda^{\nu'}B^{\nu'sij}D_j(v^s_i-G^{si}),
\]
which, by integrating by parts, entails
\[
D_i\tilde F^i=\gamma^\nu L^\nu+\alpha^{si}(v^s_i-G^{si}).
\]
Here $\alpha^{si}=-D_jB^{jis\nu'*}\lambda^{\nu'}$ and 
$\gamma^\nu=\lambda^\nu+A^{\nu\nu'*}\lambda^{\nu'}$ are functions of~$x$ and derivatives of~$u$, and
$A^{\nu\nu'*}$ and $B^{jis\nu'*}$ denotes the formally adjoint operator to $A^{\nu'\nu}$ and $B^{\nu'sij}$, respectively. 
$F$ and $\tilde F$ are equivalent conserved vectors as their difference vanishes on $\mathcal L_{\rm p}$.

Thus we obtain the characteristic $(\alpha^{si},\gamma^\nu,\,s=1,\dots,p,\,i=1,\dots,n,\,\nu\in\mathcal N)$ 
of the conservation law with conserved vector $F$,  depending exclusively on~$x$ and derivatives of~$u$.
\end{proof}

Thus we may combine
Proposition~\ref{PropositionOnInducedCLSOfFoliatedSystems}, Theorem~\ref{TheoremOnInducingExtendedCharsOfFoliatedSystems} and 
Lemmas~\ref{LemmaOnCLwithLocalCVMultiDimCaseForAbelianCoverings} and~\ref{LemmaOnCLwithLocalCharMultiDimCaseForAbelianCoverings} 
into the following result. 

\begin{theorem}\label{TheoremOnCLsOfAbelianCoverings}
The following statements on a conservation law of a system determining an Abelian covering are equivalent:

1) the conservation law is induced by a conservation law of the corresponding initial system;

2) it contains a conserved vector which does not depend on potentials;

3) some of its extended characteristics are induced by characteristics of the initial system;

4) it possesses a characteristic not depending on potentials.
\end{theorem}

\begin{note}
The locality properties of conservation laws, listed in Theorem~\ref{TheoremOnCLsOfAbelianCoverings}, 
are preserved under equivalence transformations of potential systems.
More precisely, if the systems~$\mathcal L_{\rm p}$ and~$\tilde{\mathcal L}_{\rm p}$ 
belong to the same Abelian covering of the system~$\mathcal L$ 
then the corresponding equivalence transformation maps any conservation law of~$\mathcal L_{\rm p}$ 
with these locality properties to a conservation law of~$\tilde{\mathcal L}_{\rm p}$ with the same properties. 
Therefore the statement on locality properties of conservation laws of potential systems 
can be reformulated as an analogous statement for Abelian coverings.
\end{note}


\section{Standard potentials}\label{SectionOnStandardPotsInMultiDimCase}

Consider potential systems obtained via inducing potentials according to Theorem~\ref{TheoremOnNullDivergences} in the case $n>2$.
Suppose that the system~$\mathcal L$ has $p$ linearly independent local conservation laws with conserved vectors 
$G^s=(G^{s1},\dots,G^{sn})$, $s=1,\dots,p$. 
We introduce the potentials $v^{sij}=-v^{sji}$ associated with this set of conserved vectors by the equations
\begin{equation}\label{EqPotPartInMultiDimCaseForStandardPotsForManyCLs}
v^{sij}_j=G^{si},
\end{equation}
assuming additionally that these potentials are locally independent on the solution set of the system~$\mathcal L$.
The corresponding \emph{standard potential system}~$\mathcal L_{\rm p}$ consists of 
the potential part~\eqref{EqPotPartInMultiDimCaseForStandardPotsForManyCLs} and 
the equations of the system~$\mathcal L$ which are not differential consequences 
of~\eqref{EqPotPartInMultiDimCaseForStandardPotsForManyCLs} and other equations of~$\mathcal L$, taken together. 
Below the index~$\nu$ runs through the set $\mathcal N$ of the indices of such equations 
and the index~$\nu'$ runs through the set $\mathcal N'=\{1,\dots,l\}\backslash\mathcal N$. 
(Note that the total number of such equations is equal to or greater than $l-p$ but is not necessarily equal to $l-p$.) 
The above representation is a canonical foliation of the system~$\mathcal L_{\rm p}$ over the system~$\mathcal L$. 

Tuples $v=(v^{sij})$ and $\tilde v=(\tilde v^{sij})$ of potentials 
associated with the same $p$-dimensional subspace of~the conservation law space $\CL(\mathcal L)$ of~$\mathcal L$ are considered \emph{equivalent}.
In other words, the tuples of potentials $v$ and $\tilde v$ are equivalent if 
there exist differential functions $\Phi^{sij}[u]$ and constants $c_{s\sigma}$ such that $\Phi^{sij}=-\Phi^{sji}$, $|c_{s\sigma}|\ne0$ and 
the transformation $\tilde v^{sij}=c_{s\sigma}v^{\sigma ij}+\Phi^{sij}[u]$ (the variables~$x$ and derivatives of~$u$ are not transformed) maps
the system~$\mathcal L_{\rm p}$ associated with~$v$ to the system~$\tilde{\mathcal L}_{\rm p}$ associated with~$\tilde v$. 
The tuples of the corresponding conserved vectors $G^s$ and $\tilde G^s$ 
are connected by the formula $\smash{(\tilde G^{si}-c_{s\sigma}G^{\sigma i}-D_i\Phi^{sij})\bigl|_{\mathcal L}=0}$. 
We will also say that the systems~$\mathcal L_{\rm p}$ and~$\tilde{\mathcal L}_{\rm p}$ 
are equivalent as potential systems of the system~$\mathcal L$.

The procedure of grading the jet space with respect to potentials in the case $n>2$ is analogous to the one 
in the two-dimensional case (see Section~\ref{SectionOn2DimCase}). 
The difference is that the weights of the potentials arising from the same conservation law 
(i.e., having the same value of the index~$s$) are assumed equal, i.e., 
\[
\varrho(v^{sij}_\alpha)=\max\bigl(0,\varrho(G^{s1})-1,\dots,\varrho(G^{sn})-1\bigr)+|\alpha|.
\] 

\begin{lemma}\label{LemmaOnTotalNondegeneracyOfMultiDPotSystems}
The system~$\mathcal L$ is totally nondegenerate with respect to a weight if and only if  
the system~$\mathcal L_{\rm p}$ is totally nondegenerate with respect to this weight extended to the derivatives of the potentials.
\end{lemma}

\begin{proof}
A complete set~$L_{{\rm p}[k]}$ of independent differential consequences of the system~$\mathcal L_{\rm p}$ 
which have extended weights not greater than $k$ is exhausted by the equations 
\begin{gather*}
\breve L^{\breve\mu}=0, \ \breve\mu=1,\dots,\breve l,\quad  
v^{sij}_{\alpha+\delta_j}=D_1^{\alpha_1}\dots D_n^{\alpha_n}G^{si}.
\end{gather*}
Here the equations $\smash{\breve L^{\breve\mu}=0}$, $\smash{\breve\mu=1,\dots,\breve l}$, 
form a complete set~$L_{[k]}$ of independent differential consequences of the system~$\mathcal L$, 
which have weights not greater than~$k$ and
$v^s_\alpha=\p^{|\alpha|}v^s/\p x_1^{\alpha_1}\dots\p x_n^{\alpha_n}$. 
For each~$i$ and $s$ the multiindex $\alpha=(\alpha_1,\dots,\alpha_n)$ runs through the multiindex set 
in which $\varrho(v^s)+|\alpha|<k$ and additionally $\alpha_1=0$ if $i=1$. 
The symbol $\delta_i$ was introduced after Definition~\ref{def.conservation.law}.

It is obvious that for any $k\in\mathbb N$ the system~$L_{{\rm p}[k]}$ is of maximal rank on the manifold~$\mathcal L_{{\rm p}[k]}$ 
in the weighted jet space $J^k_\varrho(x|u,v)$ if and only if the system~$L_{[k]}$ is of maximal rank on the manifold~$\mathcal L_{[k]}$. 
The local solvability of~$\mathcal L_{\rm p}$ follows from the local solvability of~$\mathcal L$ 
and the compatibility conditions for the potential part and implies the local solvability of~$\mathcal L$ 
since~$\mathcal L$ is a subsystem of~$\mathcal L_{\rm p}$. 
\end{proof}

Similarly to two-dimensional potential systems and systems representing Abelian coverings, 
multi-dimensional potential systems are foliated over the corresponding initial systems in a special way. 
In addition to using all statements of Section~\ref{SectionOnFoliatedSystemsOfDifferentialEquations},
this allows us to prove stronger statements on their conservation laws induced by conservation laws of the initial systems. 

\begin{lemma}\label{LemmaOnCLwithLocalCVMultiDimCaseForStandardPots}
If a characteristic of the potential system~$\mathcal L_{\rm p}$ depends only on local variables 
(i.e., independent and non-potential dependent ones) 
then the associated conservation law of~$\mathcal L_{\rm p}$ has a conserved vector which also does not depend on potentials. 
\end{lemma}

\begin{proof}
By assumption, the potential system~$\mathcal L_{\rm p}$ has a characteristic 
\[
(\alpha^{si},\gamma^\nu,\,s=1,\dots,p,\,i=1,\dots,n,\,\nu\in\mathcal N)
\] 
which does not depend on the potentials $v^1$,~\dots, $v^p$.  
(Here the dependence of the characteristic on nonzero derivatives of the potentials 
can be neglected up to the equivalence relation of characteristics
by~\eqref{EqPotPartInMultiDimCaseForStandardPotsForManyCLs}.) 
Since the $\alpha^{si}$ and $\gamma^\nu$ are functions of~$x$ and derivatives of~$u$ 
corresponding to the equations $D_jv^{sij}=G^{si}$ and $L^\nu=0$, respectively,
there exists a conserved vector $F$ of the potential system~$\mathcal L_{\rm p}$ with
\begin{equation}\label{EqCharFormOfPotCLMultiDimCaseForStandardPotsA}
D_iF^i=\alpha^{si}(v^{sij}_j-G^{si})+\gamma^\nu L^\nu=:V.
\end{equation}
It follows that the differential function $V=V[u,v]$ is a total divergence, so 
the extended Euler operator~$\Eop=(\Eop_{u^1},\ldots, \Eop_{u^m},\Eop_{v^{1ij}},\ldots, \Eop_{v^{pij}},\,1\leqslant i<j\leqslant n)$ 
annihilates $V$. Thus, 
\[
-\Eop_{v^{sij}}V=D_j\alpha^{si}-D_i\alpha^{sj}=0. 
\]
These conditions mean that for each~$s$ the `horizontal' differential 1-form $\omega^s=\alpha^{si}[u]\,dx_i$ 
is closed with respect to the total differential $\sf D$ since 
\[
{\sf D}\,\omega^s=D_j\alpha^{si}\,dx_j\wedge dx_i=\sum_{i<j}(D_j\alpha^{si}-D_i\alpha^{sj})dx_j\wedge dx_i=0.
\]
The `horizontal' de Rahm complex~\cite{Bocharov&Co1997} (also called $\sf D$-complex~\cite{Olver1993})
over a totally star-shaped domain of the independent variable~$x$ and the dependent variable~$u$ is exact
(see, e.g., Theorem~5.59 of~\cite{Olver1993}). 
Therefore, the form $\omega^s$ is  $\sf D$-exact, i.e., 
there exists a `horizontal' differential 0-form (in other words, a differential function) $\Phi^s=\Phi^s[u]$ such that $\omega^s={\sf D}\Phi^s$. 
Writing the last equality in components, we obtain $\alpha^{si}=D_i\Phi^s$.

Consider the conserved vector~$\hat F$ with the components 
\[
\hat F^i=F^i-\Phi^s(v^{sij}_j-G^{si})
\]
which is equivalent to the initial conserved vector $F$.
Then equation~\eqref{EqCharFormOfPotCLMultiDimCaseForStandardPotsA} can be re-written as
\[
D_i\hat F^i=-\Phi^s(v^{sij}_{ij}-D_iG^{si})+\gamma^\nu L^\nu=\Phi^sD_iG^{si}+\gamma^\nu L^\nu.
\] 
The right-hand side of this equation is a function of $x$ and derivatives of $u$ 
and vanishes on the manifold~$\mathcal L_{(k)}$ in the jet space~$J^k(x|u)$, 
where $k$ is the highest order of derivatives in this expression. 
Using the Hadamard lemma and ``integration by parts'' as in the derivation of the general characteristic form of 
conservation laws, we obtain that 
\begin{equation}\label{EqCharFormOfPotCLMultiDimCaseForStandardPotsB}
D_i\check F^i=\check\gamma^\mu L^\mu
\end{equation}
for some differential functions $\check\gamma^\mu[u]$, 
where the conserved vector $\check F$ is equivalent to $\hat F$ and, therefore, $F$ since 
it differs from $\hat F$ by a tuple vanishing on the solution set of~$\mathcal L$. 
Since the right-hand side $\check\gamma^\mu L^\mu$ depends only on~$x$ and derivatives of~$u$, 
equality~\eqref{EqCharFormOfPotCLMultiDimCaseForStandardPotsB} implies 
in view of Corollary~\ref{CorollaryOnDependenceOfConservedVectorsOnPartOfVars} that 
there exists a conserved vector $\tilde F$ of $\mathcal L_{\rm p}$, 
which depends only on~$x$ and derivatives of~$u$ and is equivalent to the conserved vector $\check F$ and, therefore, $F$.
\end{proof}

\begin{lemma}\label{LemmaOnCLwithLocalCharMultiDimCaseForStandardPots}
If an extended characteristic of the potential system~$\mathcal L_{\rm p}$ is induced by a characteristic of the system~$\mathcal L$
then the associated conservation law of~$\mathcal L_{\rm p}$ has a characteristic which does not depend on potentials. 
\end{lemma}

\begin{proof}
Assume that the multi-dimensional potential system~$\mathcal L_{\rm p}$ possesses an extended characteristic 
which is induced by a characteristic~$\lambda$ of the initial system~$\mathcal L$. This means that 
there exists a conserved vector $F=(F^1,\dots,F^n)$ of~$\mathcal L_{\rm p}$ such that $D_iF^i=\lambda^\mu[u]L^\mu[u]$.  
Again this equation need not be a characteristic form 
of the conservation law of~$\mathcal L_{\rm p}$ which contains the conserved vector $F$, 
since some equations of~$\mathcal L$ may not be contained in the minimal set of equations 
forming the potential system~$\mathcal L_{\rm p}$.
We form the set $\mathcal N'=\{\nu'\}$ of indices of such equations 
and may suppose that $\mathcal N'\ne\varnothing$. 
Then by Lemma~\ref{LemmaSimplestVersionOfHadamardLemmaForFiberBundles},
$L^{\nu'}$, being a differential consequence of~$\mathcal L_{\rm p}$, can be represented as
\[
L^{\nu'}=A^{\nu'\nu}L^\nu+B^{\nu's}D_iG^{si},
\]
where $A^{\nu'\nu}$ and $B^{\nu's}$ are polynomials of the total differentiation operators $D_i$ with 
coefficients depending on~$x$ and derivatives of~$u$.
Since $D_iG^{si}=D_i(v^{sij}_j-G^{si})$, it follows that
\[
D_iF^i=\lambda^\nu L^\nu+\lambda^{\nu'}A^{\nu'\nu}L^\nu+\lambda^{\nu'}B^{\nu's}D_i(v^{sij}_j-G^{si}),
\]
and integrating by parts on the right-hand side leads to
\[
D_i\tilde F^i=\alpha^{si}(v^{sij}_j-G^{si})+\gamma^\nu L^\nu.
\]
In this expression, $\alpha^{si}=-D_iB^{s\nu'*}\lambda^{\nu'}$ and $\gamma^\nu=\lambda^\nu+A^{\nu\nu'*}\lambda^{\nu'}$ 
are differential functions of~$u$, and
$A^{\nu\nu'*}$ and $B^{s\nu'*}$ are the formally adjoint operators to $A^{\nu'\nu}$ and $B^{\nu's}$. 
Also, $F$ and $\tilde F$ are equivalent as conserved vectors as their difference vanishes on $\mathcal L_{\rm p}$.

This gives the characteristic $(\alpha^{si},\gamma^\nu,\,s=1,\dots,p,\,i=1,\dots,n,\,\nu\in\mathcal N)$ 
of the conservation law with conserved vector $F$, which depend only on~$x$ and derivatives of~$u$, as claimed.
\end{proof}

Summarizing
Proposition~\ref{PropositionOnInducedCLSOfFoliatedSystems}, Theorem~\ref{TheoremOnInducingExtendedCharsOfFoliatedSystems} and 
Lemmas~\ref{LemmaOnCLwithLocalCVMultiDimCaseForStandardPots} and~\ref{LemmaOnCLwithLocalCharMultiDimCaseForStandardPots}, we arrive at:

\begin{theorem}\label{TheoremOnCLsOfMultiDPotSystems}
The following statements on a conservation law of a standard potential system (without gauges) are equivalent:

1) the conservation law is induced by a conservation law of the corresponding initial system;

2) it contains a conserved vector which does not depend on potentials;

3) some of its extended characteristics are induced by characteristics of the initial system;

4) it possesses a characteristic not depending on potentials.
\end{theorem}

\begin{note}
The locality properties of conservation laws, listed in Theorem~\ref{TheoremOnCLsOf2DPotSystems}, 
are stable with respect to the equivalence of potential systems.
In other words, if potential systems~$\mathcal L_{\rm p}$ and~$\tilde{\mathcal L}_{\rm p}$ of the system~$\mathcal L$ are 
equivalent then the corresponding equivalence transformation maps any conservation law of~$\mathcal L_{\rm p}$ 
possessing the above locality properties to a conservation law of~$\tilde{\mathcal L}_{\rm p}$ with the same properties. 
\end{note}

If~\mbox{$n>2$}, the equations~\eqref{EqPotPartInMultiDimCaseForStandardPotsForManyCLs} 
associated with a fixed solution $u=u(x)$ of the system~$\mathcal L$ 
form an underdetermined system with respect to the potentials~$v^{sij}$.
Therefore, we can add gauge conditions on the potentials to~$\mathcal L_{\rm p}$. 
In fact, such additional conditions are absolutely necessary in the case \mbox{$n>2$} for the potential system 
to have nontrivial symmetries and conservation laws. 
It is stated in Theorem 2.7 of~\cite{Anco&Bluman1997} for a quite general situation that 
every local symmetry of a potential system with unconstrained potentials is projectable to a local symmetry of the initial system, 
i.e., such a potential system gives no nontrivial potential symmetries. 
Moreover, each conservation law of such a system is invariant with respect to gauge transformations of the potentials \cite{Anco&The2005}.

\begin{definition}\label{DefinitionOfGaugesOnPots}
A system~$\mathcal L_{\rm g}$ of differential equations 
with the independent variables~$x$ and the dependent variables~$u$ and~$v$ is called a \emph{gauge} on 
the potentials~$v^{sij}$ defined by equations~\eqref{EqPotPartInMultiDimCaseForStandardPotsForManyCLs} 
if any differential consequence of the coupled system $\mathcal L_{\rm gp}=\mathcal L_{\rm p}\cap\mathcal L_{\rm g}$, 
which does not involve the potentials $v^{sij}$, is a differential consequence of the initial system~$\mathcal L$. 
The coupled system $\mathcal L_{\rm gp}$ is called a \emph{gauged potential system}.
The gauge~$\mathcal L_{\rm g}$ is called \emph{weak} 
if a minimal set of equations generating all the differential consequences of~$\mathcal L_{\rm p}$ 
contains a minimal set of the coupled system $\mathcal L_{\rm gp}$ called a \emph{weakly gauged potential system}. 
\end{definition}

The gauged potential system $\mathcal L_{\rm gp}$ is a foliated system over the base system~$\mathcal L$. 
Therefore, the statements of Section~\ref{SectionOnFoliatedSystemsOfDifferentialEquations} are true for conservation laws of such systems 
and can be sharpened in the following way.

\begin{proposition}\label{PropositionOnInducingCLsMultiDimCaseForStandardGaugedPots}
A conservation law of a gauged potential system contains a conserved vector which does not depend on potentials 
if and only if it is induced by the conservation law of the corresponding initial system with the same conserved vector 
and if and only if some of its extended characteristics are induced by characteristics of the initial system.
\end{proposition}

A weakened version of Theorem~\ref{TheoremOnCLsOfMultiDPotSystems} on potential systems without gauges can be extended 
to weakly gauged potential systems. 
The proof is analogous to those already presented. 
Only the general version of the Hadamard lemma for fiber bundles (Lemma~\ref{LemmaGeneralVersionOfHadamardLemmaForFiberBundles}) 
has to be applied instead of the simplest one (Lemma~\ref{LemmaSimplestVersionOfHadamardLemmaForFiberBundles}). 

\begin{theorem}\label{TheoremOnCLsOfMultiDWeaklyGaugedPotSystems}
A conservation law of a weakly gauged potential system 
contains a conserved vector which does not depend on potentials
if and only if it has a characteristic which also does not depend on potentials 
and whose components corresponding to the gauge equations vanish. 
\end{theorem}


\section{General coverings}\label{SectionOnGeneralCoverings}

The idea of general coverings arose in the well-known paper by Wahlquist and Estabrook~\cite{Wahlquist&Estabrook1975} 
in the form of prolongation structures involving \emph{pseudopotentials}. 
Later this idea was rigorously formulated and developed in geometrical terms 
\cite{Bocharov&Co1997,Krasil'shchik&Vinogradov1984,Krasil'shchik&Vinogradov1989,Vinogradov&Krasil'shchik1984}. 
Here we treat coverings in the framework of the local approach by introducing local coordinates. 

The statement on the simultaneous locality of conserved vectors and characteristics is not true for conservation laws of 
general coverings. 

Suppose that the system~$\mathcal L$ admits $p$ pseudo-potentials $v^1$,~\dots, $v^p$ defined by the equations 
\begin{equation}\label{EqPotPartInMultiDimCaseForGeneralCoverings}
v^s_i=G^{si}[u|v], 
\end{equation}
where differential functions $G^{si}=G^{si}[u|v]$ 
satisfy the compatibility conditions $\hat D_jG^{si}=\hat D_iG^{sj}$ on the solution set of the system~$\mathcal L$. 
The notation $G[u|v]$ means that $G$ is a differential function of $u$ and $v$, depending on~$x$, $v$ and derivatives of~$u$ 
(there are no derivatives of~$v$ of orders greater than~0!). 
We will briefly call $G[u|v]$ a differential function of $(u|v)$.
$\hat D_i$ is the operator of total differentiation, acting on differential functions of the above type according to 
system~\eqref{EqPotPartInMultiDimCaseForGeneralCoverings}, i.e., 
$\hat D_i=\p_{x_i}+u^a_{\alpha,i}\p_{u^a_\alpha}+G^{si}[u|v]\p_{v^s}$.

The corresponding potential system~$\mathcal L_{\rm p}$ consists of the pseudo-potential part~\eqref{EqPotPartInMultiDimCaseForGeneralCoverings} 
and the equations of the system~$\mathcal L$ which are not differential consequences of \eqref{EqPotPartInMultiDimCaseForGeneralCoverings} 
together with other equations of~$\mathcal L$. 
The system~$\mathcal L_{\rm p}$ defines a \emph{covering} of the system~$\mathcal L$. 
It is an example of a foliated system, where $\mathcal L$ is the base system.

Two tuples of pseudo-potentials $v=(v^1,\dots,v^p)$ and $\tilde v=(\tilde v^1,\dots,\tilde v^p)$ 
of the same system~$\mathcal L$ are considered \emph{equivalent} 
if there exist differential functions $\Omega^s[u|v]$ such that $|\Omega^s_{v^\sigma}|\ne0$ and if
the transformation~$\Omega$: $\tilde v^s=\Omega^s[u|v]$ (the variables~$x$ and derivatives of~$u$ are not transformed) maps
the system~$\mathcal L_{\rm p}$ associated with~$v$ to the system~$\tilde{\mathcal L}_{\rm p}$ associated with~$\tilde v$. 
The functions $G^{si}[u|v]$ and $\tilde G^{si}[u|\tilde v]$ from the pseudo-potential parts of these systems 
are connected by the formula $(\tilde G^{si}-\hat D_i\Omega^s)\bigl|_{\mathcal L}=0$. 
Hence the prolongations of the total differentiation operators to equivalent tuples of pseudo-potentials 
coincide on the solution set of~$\mathcal L$.
In fact, in the local-coordinate approach a covering of~$\mathcal L$ is 
an equivalence class of tuples of pseudo-potentials which are considered along with 
the corresponding equations of the form~\eqref{EqPotPartInMultiDimCaseForGeneralCoverings} 
and prolongations of the total differentiation operators coinciding on the solution set of~$\mathcal L$.

Since two conserved vectors of~$\mathcal L_{\rm p}$, 
whose difference vanises identically in view of subsystem~\eqref{EqPotPartInMultiDimCaseForGeneralCoverings} are equivalent, 
any conservation law of~$\mathcal L_{\rm p}$ contains a conserved vector $F[u|v]$ whose components $F^i[u|v]$
do not depend on nonzero-order derivatives of the pseudo-potentials. 
In view of Lemma~\ref{LemmaSimplestVersionOfHadamardLemmaForFiberBundles},
the defining formula $\smash{D_iF^i\bigl|_{\mathcal L_{\rm p}}=0}$ for conserved vectors of this kind can be rewritten in the form 
$\smash{\hat D_iF^i\bigl|_{\mathcal L}=0}$. 
The same is true for characteristics and extended characteristics of the system~$\mathcal L_{\rm p}$. 
Namely, up to equivalence determined by the subsystem~\eqref{EqPotPartInMultiDimCaseForGeneralCoverings}, 
the components of any (extended) characteristic of~$\mathcal L_{\rm p}$ can be assumed to be differential functions of $(u|v)$. 
Conserved vectors (characteristics and extended characteristics) 
whose components do not depend on the nonzero-order derivatives of the pseudo-potentials will be called \emph{reduced}.

Due to the structure of the equations \eqref{EqPotPartInMultiDimCaseForGeneralCoverings} defining the pseudo-potentials, 
any weight defined for the variables~$x$ and~$u^a_\alpha$ is extendable to the derivatives of pseudo-potentials.
To extend the weight, we use the following rule: 
We will assume that all the pseudo-potentials~$v$ have the same weight equal, e.g., to 
\[
\varrho_v=\max\bigl(0,\varrho({G^si})-1,\,s=1,\dots,p,\,i=1,\dots,n\bigr).
\]
Therefore, $\varrho(v^s_\alpha)=\varrho_v+|\alpha|$. 
This equation reflects the fact that pseudo-potentials appear 
on the right-hand sides of the equations~\eqref{EqPotPartInMultiDimCaseForGeneralCoverings}.

The jet spaces can also be endowed with weights with respect to pseudo-potentials by means of the same rule as for 
the usual potentials in the two-dimensional case. See Section~\ref{SectionOn2DimCase} for notations and definitions. 

\begin{lemma}\label{LemmaOnTotalNondegeneracyOfGeneralCoverigs}
The system~$\mathcal L$ is totally nondegenerate with respect to a weight if and only if  
the system~$\mathcal L_{\rm p}$ is totally nondegenerate with respect to this weight extended to the derivatives of the pseudo-potentials.
\end{lemma}

The proof of Lemma~\ref{LemmaOnTotalNondegeneracyOfGeneralCoverigs} is analogous to that of 
Lemma~\ref{LemmaOnTotalNondegeneracyOfAbelianCoverigs}. 
Only the total differentiation operators $\hat D_i$ have to be used instead of the standard ones. 
Thus only the total nondegeneracy of the system~$\mathcal L$ has to be assumed for working with 
the usual and extended characteristics of conservation laws of both the system~$\mathcal L$ and the system~$\mathcal L_{\rm p}$.
Since any potential system determining a covering of the system~$\mathcal L$ is a foliated system with base system~$\mathcal L$, 
the statements of Section~\ref{SectionOnFoliatedSystemsOfDifferentialEquations} remain true for conservation laws of such systems 
(after the necessary replacements in the proof of Theorem~\ref{TheoremOnInducingExtendedCharsOfFoliatedSystems}, 
taking into account the grading of the jet spaces).  
Let us combine these statements and formulate them in a specific way.

\begin{proposition}\label{PropositionOnInducingCLsMultiDimCaseForGeneralCoverings}
A conservation law of a system determining a covering contains a conserved vector which does not depend on potentials 
if and only if it is induced by the conservation law of the corresponding initial system which has the same conserved vector 
and if and only some of its extended characteristics are induced by characteristics of the initial system.
\end{proposition}

Unfortunately, the property of characteristic locality cannot be included in the chain of equivalent statements of 
Proposition~\ref{PropositionOnInducingCLsMultiDimCaseForGeneralCoverings} 
and, moreover, this property is not preserved under the equivalence transformations of tuples of pseudo-potentials. 
In fact, if the potential systems~$\mathcal L_{\rm p}$ and~$\tilde{\mathcal L}_{\rm p}$ of the system~$\mathcal L$ are 
equivalent with respect to an equivalence transformation~$\Omega$ and 
the system~$\mathcal L_{\rm p}$ possesses a conservation law~$\mathcal F$ with a local characteristic 
associated with equivalent tuples of pseudo-potentials 
then there is no guarantee that the conservation law~$\tilde{\mathcal F}$ of~$\tilde{\mathcal L}_{\rm p}$, 
equivalent to~$\mathcal F$ with respect to~$\Omega$, also has a local characteristic. 

A partial locality property of extended characteristics of covering systems is connected with the linearity of 
associated conserved vectors with respect to pseudo-potentials.

\begin{theorem}\label{TheoremOnCLsOfGeneralCoverings}
A conservation law of a system determining a covering contains a reduced conserved vector which linearly depends on pseudo-potentials
if and only if 
it has a reduced extended characteristic whose components corresponding to the pseudo-potential part of the system 
do not depend on pseudo-potentials.
\end{theorem}

\begin{proof}
Suppose that a conservation law~$\mathcal F$ of the system~$\mathcal L_{\rm p}$ contains a reduced conserved vector $F[u|v]$ 
which linearly depends on pseudo-potentials, i.e., $F^i=F^{is}[u]v^s+F^{i0}[u]$.
The defining formula $\smash{\hat D_iF^i\bigl|_{\mathcal L}=0}$ for reduced conserved vectors implies that 
\[
\bigl((D_iF^{is})v^s+F^{is}G^{si}+D_iF^{i0}\bigr)\bigl|_{\mathcal L}=0.
\]
Following the conventional way of deriving the characteristic form of conservation laws, 
we apply the Hadamard lemma, integrate by parts on the right-hand side of the derived equality and finally obtain that 
\[
(D_iF^{is})v^s+F^{is}G^{si}+D_iF^{i0}=\gamma^\mu L^\mu+D_i\hat F^i
\]
for some differential functions $\gamma^\mu=\gamma^\mu[u|v]$ and $\hat F^i=\hat F^i[u|v]$,
and the functions $\hat F^i$ vanish on the solutions of~$\mathcal L$ identically with respect to~$v$.
Therefore, the tuple $\hat F=(\hat F^1,\dots,\hat F^n)$ is a trivial conserved vector of~$\mathcal L_{\rm p}$. 
The conserved vector $\tilde F=F-\hat F$ belongs to~$\mathcal F$ (since it is equivalent to~$F$) and satisfies the equality 
\[
D_i\tilde F^i=F^{is}(v^s_i-G^{si})+\gamma^\mu L^\mu.
\]
This means that the tuple $(F^{is}[u], i=1,\dots,n, s=1,\dots,p, \gamma^\mu[u|v], \mu=1,\dots,l)$ is a reduced extended characteristic 
of the system~$\mathcal L_{\rm p}$, which is associated with the conservation law~$\mathcal F$ and obviously has the necessary property.

Conversely, let the tuple $(F^{is}[u], i=1,\dots,n, s=1,\dots,p, \gamma^\mu[u|v], \mu=1,\dots,l)$ be a reduced extended characteristic 
associated with the conservation law~$\mathcal F$ of the system~$\mathcal L_{\rm p}$. 
Then there exists a conserved vector~$F$ belonging to~$\mathcal F$ such that 
\begin{equation}\label{EqExtendedCharFormOfCLsForGeneralCoverings}
D_iF^i=F^{is}(v^s_i-G^{si})+\gamma^\mu L^\mu.
\end{equation}
Acting by the extended Euler operator~$\Eop=(\Eop_{u^1},\ldots, \Eop_{u^m},\Eop_{v^1},\ldots, \Eop_{v^p})$ 
on both the sides of the last equality, we have in particular that 
\[
0=\Eop_{v^s}D_iF^i= -D_iF^{is}-F^{is}G^{si}_{v^s}+\gamma^\mu_{v^s} L^\mu. 
\]
Simultaneously integrating these equations, we obtain that 
\[
-F^{is}G^{si}+\gamma^\mu L^\mu=(D_iF^{is})v^s+H[u]
\]
for some differential function~$H=H[u]$. 
The substitution of the last expression into equation~\eqref{EqExtendedCharFormOfCLsForGeneralCoverings} results in 
the equality $D_iF^i=F^{is}v^s_i+(D_iF^{is})v^s+H$, i.e., $D_i(F^i-F^{is}v^s)=H[u]$. 
This immediately implies in view of Corollary~\ref{CorollaryOnDependenceOfConservedVectorsOnPartOfVars} 
that there exist an $n$-tuple $\breve F=\breve F[u]$ and a null divergence $\check F=\check F[u,v]$ 
such that $F^i-F^{is}v^s=\breve F^i+\check F^i$. 
Finally, the tuple $\tilde F=F-\check F$ differs from $F$ by the null divergence $\check F$ and, therefore, 
also is a conserved vector of~$\mathcal L_{\rm p}$, belonging to the conservation law~$\mathcal F$. 
Its components $\tilde F^i=F^{is}[u]v^s+\breve F^i[u]$ are linear with respect to the pseudo-potentials. 
\end{proof}


\section{A criterion for purely potential conservation laws}\label{SectionOnCriterionForPurelyPotentialCLs}

The main applications of the results collected in Theorem~\ref{TheoremUnitedOnCLsOfDifferentPotSystems} 
are connected with the construction of potential (nonlocal) conservation laws and hierarchies of potential systems. 
At first sight it appears that they are important mostly for those approaches to finding conservation laws 
which involve the characteristic form~\eqref{EqCharFormOfConsLaw} of conservation laws or its consequences~\eqref{EqNSCondOnChar} 
and~\eqref{EqNCondOnChar}, including the Noether symmetry approach \cite{Anco&Bluman2002a,Anco&Bluman2002b,Bocharov&Co1997,Olver1993}. 
(A detailed comparative analysis of different methods of finding conservation laws and their realizations is given in~\cite{Wolf2002}.) 
A more careful consideration reveals that these results are also important for the direct method based on the definitions 
of conserved vectors and conservation laws~\cite{Popovych&Ivanova2004CLsOfNDCEs}. 
Given a conserved vector depending on derivatives of potentials, usually it is difficult to test 
whether this conserved vector is equivalent to a conserved vector which does not depend on potentials. 
The reason of the difficulty is the duplicity of the equivalence relation of conserved vectors, 
which is generated by summands of two kinds---null divergences and tuples of differential functions identically vanishing on 
the solution set of the corresponding system of differential equations. 
That is why it seems impossible to formulate, directly in terms of conserved vectors, an effective criterion 
for testing whether a conservation law of a potential system is induced by a conservation law of the corresponding initial system.
At the same time, such a criterion is easily formulated  in terms of characteristics. 

\begin{proposition}\label{PropositionOnEssentialChars}
Let a system~$\mathcal L$ be totally nondegenerate with respect to a weight, 
$\mathcal L_{\rm p}$ be a system determining an Abelian covering of~$\mathcal L$ 
(resp. a potential system of~$\mathcal L$ in the two-dimensional case). Moreover, let  
a characteristic~$\lambda$ of $\mathcal L_{\rm p}$ be completely reduced, i.e., 
the derivatives of potentials of orders greater than 0 are excluded from~$\lambda$ due to differential consequences of the potential part 
of~$\mathcal L_{\rm p}$ and then the constrained derivatives of~$u$ are excluded from~$\lambda$ due to differential consequences of~$\mathcal L$. 
Then the characteristic~$\lambda$ is associated with a conservation law of~$\mathcal L_{\rm p}$, which is not induced 
by a conservation law of~$\mathcal L$, if and only if it depends on potentials.  
\end{proposition}

\begin{proof}
If a characteristic~$\lambda$ of $\mathcal L_{\rm p}$ is completely reduced and depends on potentials 
then it is unconditionally inequivalent to any characteristic free from all derivatives of potentials.  
That is why the necessary statement directly follows from Theorem~\ref{TheoremOnCLsOfAbelianCoverings}
(resp. Theorem~\ref{TheoremOnCLsOf2DPotSystems}).
\end{proof}

Let us consider the two-dimensional case in some more detail, employing the notations of Section~\ref{SectionOn2DimCase}.
Suppose that a conserved vector $(F,G)$ of a potential system~$\mathcal L_{\rm p}$ is associated with  
a characteristic 
\[
\lambda=(\alpha^s[u],\beta^s[u],\gamma^\nu[u],\,s=1,\dots,p,\,\nu\in\mathcal N)
\]
which does not depend on derivatives of potentials.
Then we can algorithmically find a conserved vector~$(\tilde F,\tilde G)$ 
which is equivalent to~$(F,G)$ and also does not depend on derivatives of potentials, 
avoiding the direct application of the complicated formula from Theorem~\ref{TheoremOnEqsWithDivergence}. 
The algorithm is based on the proof of Lemma~\ref{LemmaOnCLwithLocalCV2Dcase}.
Since each tuple $(\alpha^s,\beta^s)$ is a null divergence, 
there exist differential functions $\Phi^s[u]$ such that $D_x\Phi^s=\alpha^s$ and $D_t\Phi^s=-\beta^s$.
Then the conserved vector with the components  
\[
\hat F=F+\Phi^s(v^s_x-F^s), \quad \hat G=G-\Phi^s(v^s_t+G^s).
\]
is equivalent to the initial conserved vector $(F,G)$ since the difference of $(F,G)$ and $(\hat F,\hat G)$ vanishes on 
the solution set of~$\mathcal L_{\rm p}$, and the total divergence of $(\hat F,\hat G)$ is a differential function of~$u$. 
This means that the conserved vector~$(\tilde F,\tilde G)$ differs from  $(\hat F,\hat G)$ by a null divergence whose components 
are, in general, differential functions of~$u$ and~$v$. 

Suppose that the potential system $\mathcal L_{\rm p}$ has $q$ linearly independent conservation laws 
induced by conservation laws of the initial system~$\mathcal L$. 
Let the tuples $(\tilde F^\varsigma,\tilde G^\varsigma)$, $\varsigma=1,\dots,q$, be conserved vectors of 
these conservation laws which do not depend on derivatives of potentials. 
The second-level potential system (see~\cite{Popovych&Ivanova2004CLsOfNDCEs} for definitions) 
constructed from~$\mathcal L_{\rm p}$ with the conserved vectors $(\tilde F^\varsigma,\tilde G^\varsigma)$, $\varsigma=1,\dots,q$, 
is equivalent, with respect to a local transformation changing only potentials, 
to the first-level potential system $\mathcal L_{\rm p}'$ obtained from~$\mathcal L$ with the conserved vectors 
$(F^s,G^s)$, $s=1,\dots,p$, and $(\tilde F^\varsigma,\tilde G^\varsigma)$, $\varsigma=1,\dots,q$,
(cf. the end of Section~\ref{SectionOnBasicDefsAndStatementsOnCLs}). 
The potential part of $\mathcal L_{\rm p}'$ differs from the potential part of $\mathcal L_{\rm p}$ in the equations 
$v^{p+\varsigma}_x=\tilde F^\varsigma$, $v^{p+\varsigma}_t=-\tilde G^\varsigma$, $\varsigma=1,\dots,q$. 
An analogous argument holds for potential systems of an arbitrary level.

\section{An example}\label{SectionExample}

To present an illustrative example, we give a new detailed interpretation of results from~\cite{Popovych&Ivanova2004CLsOfNDCEs} 
on hierarchies of conservation laws and potential systems of diffusion--convection equations, involving tools developed in this paper. 
See also \cite{Ivanova&Popovych&Sophocleous2004,Popovych&Ivanova2004CLsOfNDCEs,Popovych&Kunzinger&Ivanova2008} 
for the method of classification of potential conservation laws for a class of differential equations 
with respect to the equivalence group of this class.

The class of diffusion--convection equations of the general form 
\begin{equation} \label{eqf1}
u_t=(A(u)u_x)_x+B(u)u_x,
\end{equation}
where $A=A(u)$ and $B=B(u)$ are arbitrary smooth functions of $u$, $A\neq0$, possesses the equivalence group $G^\sim$
formed by the transformations
\[
\tilde t=\varepsilon_4t+\varepsilon_1, \quad
\tilde x=\varepsilon_5x+\varepsilon_7 t+\varepsilon_2, \quad
\tilde u=\varepsilon_6u+\varepsilon_3, \quad
\tilde A=\varepsilon_4^{-1}\varepsilon_5^2A, \quad
\tilde B=\varepsilon_4^{-1}\varepsilon_5B-\varepsilon_7,
\]
where $\varepsilon_1,$ \dots, $\varepsilon_7$ are arbitrary constants,$\varepsilon_4\varepsilon_5\varepsilon_6\ne0.$
The kernel (intersection) $G^\cap$ of the maximal Lie invariance groups of equations from class~\eqref{eqf1} consists of the transformations
$\tilde t=t+\varepsilon_1$, $\tilde x=x+\varepsilon_2$, $\tilde u=u$.
 
Any equation from class~\eqref{eqf1} has the conservation law $\mathcal F^0$ 
whose density, flux and characteristic are 
\[\textstyle
\mathcal F^0=\mathcal F^0(A,B)\colon\quad F=u,\quad G=-Au_x-\int\! B,\quad \lambda=1.
\]
A complete list of $G^\sim$-inequivalent equations~\eqref{eqf1} having
additional (i.e., linearly independent of~$\mathcal F^0$) conservation laws
is exhausted by the following ones
\[\arraycolsep=0ex\begin{array}{lllll}
B=0,\quad&\mathcal F^1=\mathcal F^1(A)\colon\quad& F=xu,& \quad G=\int\! A-xAu_x,&\quad \lambda=x;\\[1ex]
B=A,\quad&\mathcal F^2=\mathcal F^2(A)\colon\quad& F=e^xu,& \quad G=-e^xAu_x,&\quad \lambda=e^x;\\[1ex]
A=1,\quad& B=0,\quad\mathcal F^3_h\colon\quad&  F=h u,& \quad G=h_xu-h u_x,&\quad \lambda=h.
\end{array}\]
where $\int\! A=\int\! A(u) du$, $\int\! B=\int\! B(u) du$,
$h=h(t,x)$ is an arbitrary solution of the backward linear heat equation $h_t+h_{xx}=0$.
(Along with constrains for $A$ and $B$ the above table also contains complete lists of
densities, fluxes and characteristics of additional conservation laws.)

\medskip

{\noindent\bf General case.} 
In the general case equation~\eqref{eqf1} has the unique linearly independent local conservation law $\mathcal F^0(A,B)$.
The corresponding potential system 
\[\textstyle v^1_x=u,\quad v^1_t=Au_x+\int\! B\] 
possesses only the zero conservation law, 
i.e., equation~\eqref{eqf1} of the general form admits no purely potential conservation laws.

\medskip

{\noindent\mathversion{bold}$B=0$.} 
Any equation with $B=0$ and a general value of~$A$ admits exactly 
two linearly independent local conservation laws $\mathcal F^0=\mathcal F^0(A,0)$ and $\mathcal F^1=\mathcal F^1(A)$,
and up to linear dependence any conservation law is $G^\cap$-equivalent to one of them.
Using these conservation laws, we introduce the potentials $v^1$ and $v^2$, where
\begin{gather}
v^1_x=u,\quad v^1_t=Au_x, \label{potsysB0gen}\\[1ex]
\textstyle  v^2_x=xu,\quad v^2_t=xAu_x-\int\!A.  \label{potsysB0spec}
\end{gather}
The pairs of equations~\eqref{potsysB0gen} and~\eqref{potsysB0spec}, considered separately, form two potential systems
for equation~\eqref{eqf1} (with vanishing $B$) in the unknown functions $(u,v^1)$ and $(u,v^2)$, respectively. 
The third potential system is formed by~\eqref{potsysB0gen} and~\eqref{potsysB0spec} simultaneously, 
and the three functions $u$, $v^1$ and $v^2$ are assumed unknown. 
Since the characteristics $\lambda=1$ and $\lambda=x$ are nonsingular,
the initial equation is a differential consequence of both the potential parts~\eqref{potsysB0gen} and~\eqref{potsysB0spec} 
and is not included in the minimal sets of equations representing the potential systems. 
Therefore, the characteristics of systems~\eqref{potsysB0gen} and~\eqref{potsysB0spec} have two components. 
The components~$\beta$ and $\alpha$ correspond to the first and second equations of these systems, respectively. 

System~\eqref{potsysB0gen} has only one linearly independent local conservation law~$\mathcal F$
whose conserved vector $(F,G)=(v^1,-\int\!A)$ is associated with the characteristic $(\alpha,\beta)=(1,0)$. 
In view of Theorem~\ref{TheoremOnCLsOf2DPotSystems}, this conservation law is induced by a conservation law of 
the initial equation. 
Let us find a conserved vector~$(\tilde F,\tilde G)$ 
which is equivalent to~$(F,G)$ and additionally does not depend on derivatives of potentials.
The function~$\Phi$ (see Section~\ref{SectionOnCriterionForPurelyPotentialCLs}) satisfies the equations $D_x\Phi=\alpha=1$ and $D_t\Phi=-\beta=0$. 
We choose the value $\Phi=x$ and consider the conserved vector $(\hat F,\hat G)$ equivalent to~$(F,G)$ with the components 
\begin{gather*}
\hat F=F+\Phi(v^1_x-u)=v^1+x(v^1_x-u)=(xv^1)_x-xu, \\ \textstyle
\hat G=G-\Phi(v^1_t-Au_x)=-\int\!A-x(v^1_t-Au_x)=-(xv^1)_t-\int\!A+xAu_x.
\end{gather*}
Up to the summand $((xv^1)_x,-(xv^1)_t)$ which obviously is a null divergence, 
the conserved vector $(\hat F,\hat G)$ is equivalent to the conserved vector $(\tilde F,\tilde G)=(-xu,xAu_x-\int\!A)$ 
belonging to the conservation law $-\mathcal F^1$. 
That is why the ``second-level'' potential system 
\begin{equation}\label{potsys2B0gen}
\textstyle
v^1_x=u,\quad w^1_x=v^1,\quad w^1_t=\int\!A
\end{equation}
obtained from~\eqref{potsysB0gen} by introducing the ``second-level'' potential $w^1$ with the conservation law~$\mathcal F$ 
is in fact equivalent, with respect to the point transformation $w^1=xv^1-v^2$, 
to the ``first-level'' united potential system~\eqref{potsysB0gen}--\eqref{potsysB0spec}. 
Although system~\eqref{potsys2B0gen} formally belongs to the second level,
it is the most convenient one for further investigation since it has the simplest form 
among the potential systems constructed with two conservation laws from equation~\eqref{eqf1} with $B=0$.

Analogously, system~\eqref{potsysB0spec} possesses only one linearly independent local conservation law~$\mathcal F$
with the conserved vector $(F,G)=(x^{-2}v^2,-x^{-1}\int\!A)$ and the characteristic $(\alpha,\beta)=(x^{-2},0)$. 
Theorem~\ref{TheoremOnCLsOf2DPotSystems} implies that this conservation law is induced by a conservation law of 
the initial equation. 
As a solution of the equations $D_x\Phi=\alpha=x^{-2}$ and $D_t\Phi=-\beta=0$, we choose the value $\Phi=-x^{-1}$.
Then 
\begin{gather*}
\hat F=x^{-2}v^2-x^{-1}(v^2_x-xu)=-(x^{-1}v^2)_x+u, \\ \textstyle
\hat G=-x^{-1}\int\!A+x^{-1}(v^2_t-xAu_x+\int\!A)=-(x^{-1}v^2)_t-Au_x.
\end{gather*}
The conserved vector $(\hat F,\hat G)$ is equivalent, by construction, to~$(F,G)$ on the solution set of~\eqref{potsys2B0gen}. 
Up to the null divergence $((x^{-1}v^2)_x,-(x^{-1}v^2)_t)$, it is also equivalent to the conserved vector $(\tilde F,\tilde G)=(u,-Au_x)$ 
which depends only on derivatives of~$u$ and belongs to the conservation law $\mathcal F^0$. 
Therefore the ``second-level'' potential system 
\[\textstyle
v^2_x=xu,\quad w^2_x=x^{-2}v^2,\quad w^2_t=x^{-1}\int\!A
\]
obtained from~\eqref{potsysB0spec} by introducing the ``second-level'' potential $w^2$ with the conservation law~$\mathcal F$ 
is also equivalent, with respect to the point transformation $w^2=v^1-x^{-1}v^2$, 
to the united system~\eqref{potsysB0gen}--\eqref{potsysB0spec}. 

The space of conservation laws of the united system~\eqref{potsysB0gen}--\eqref{potsysB0spec} is zero-dimensional. 
Therefore, for any equation~\eqref{eqf1} with $B=0$ all potential conservation laws are induced by local ones 
and all inequivalent potential systems are exhausted by systems~\eqref{potsysB0gen}, \eqref{potsysB0spec} and~\eqref{potsys2B0gen}. 

\medskip

{\noindent\mathversion{bold}$B=A$.}
This case is analyzed in a way similar to the previous one.
Any equation with $B=A$ and a general value of~$A$ has a two-dimensional space of local conservation laws
generated by $\mathcal F^0=\mathcal F^0(A,A)$ and $\mathcal F^2=\mathcal F^2(A)$,
and up to linear dependence any conservation law is $G^\cap$-equivalent to 
either $\mathcal F^0$ or $\mathcal F^2+\varepsilon\mathcal F^0$, 
where $\varepsilon\in\{0,\pm1\}\bmod G^\cap$.
Using the conservation laws $\mathcal F^0$ and $\mathcal F^2+\varepsilon\mathcal F^0$, we can introduce the 
independent potentials $v^1$ and $v^3$, satisfying the conditions
\begin{gather}\textstyle\label{potsysBAgen}
v^1_x=u,\quad v^1_t=Au_x+\int\!A, 
\\[1ex] \textstyle\label{potsysBAspec}
v^3_x=(e^x+\varepsilon)u,\quad v^3_t=(e^x+\varepsilon)Au_x+\varepsilon \int\!A.  
\end{gather}
The pairs of equations~\eqref{potsysBAgen} and~\eqref{potsysBAspec} considered separately form two potential systems
for equation~\eqref{eqf1} with $B=A$ in the unknown functions $(u,v^1)$ and $(u,v^3)$, respectively. 
The third potential system is formed by equations~\eqref{potsysBAgen} and~\eqref{potsysBAspec}
simultaneously, and the three functions $u$, $v^1$ and $v^3$ are assumed as unknown.
Since the characteristics $\lambda=1$ and $\lambda=e^x+\varepsilon$ are nonsingular,
the initial equation is a differential consequence of both the potential parts~\eqref{potsysBAgen} and~\eqref{potsysBAspec}
and is not included in the minimal sets of equations representing the potential systems. 
Therefore, characteristics of systems~\eqref{potsysBAgen} and~\eqref{potsysBAspec} have two components. 
The components~$\beta$ and $\alpha$ correspond to the first and second equations of these systems, respectively. 

System~\eqref{potsysBAgen} has only one linearly independent local conservation law~$\mathcal F$
whose conserved vector $(F,G)=(e^xv^1,-e^x\int\!A)$ is associated with the characteristic $(\alpha,\beta)=(e^x,0)$. 
We choose the solution $\Phi=e^x$ of the equations $D_x\Phi=\alpha=1$ and $D_t\Phi=-\beta=0$ and put 
\begin{gather*}
\hat F=e^xv^1+e^x(v^1_x-u)=(e^xv^1)_x-e^xu, \\ \textstyle
\hat G=-e^x\int\!A-e^x(v^1_t-Au_x-\int\!A)=-(e^xv^1)_t+e^xAu_x.
\end{gather*}
The conserved vector $(\hat F,\hat G)$ is equivalent to~$(F,G)$ by construction and, 
up to the null divergence $((e^xv^1)_x,-(e^xv^1)_t)$, 
is equivalent to the conserved vector $(\tilde F,\tilde G)=(-e^xu,e^xAu_x)$. This vector 
does not depend on the potential~$v^1$ and belongs to the conservation law $-\mathcal F^2$. 
Hence the conservation law~$\mathcal F$ of the potential system~\eqref{potsysBAspec} 
is induced by the conservation law $-\mathcal F^2$ of the initial equation. 
Therefore, the ``second-level'' potential system 
\begin{equation}\label{potsys2BAgen}\textstyle
v^1_x=u,\quad w^1_x=e^xv^1,\quad w^1_t=e^x\int\!A,
\end{equation}
obtained from~\eqref{potsysBAspec} by introducing the ``second-level'' potential $w^1$ with the conservation law~$\mathcal F$ 
is equivalent, with respect to the point transformation $w^1=e^xv^1-v^3$, 
to the united system~\eqref{potsysBAgen}--\eqref{potsysBAspec}, where $\varepsilon=0$. 
Although system~\eqref{potsys2BAgen} formally belongs to the second level,
it is most convenient for our further investigation among the potential systems 
constructed with two conservation laws from equation~\eqref{eqf1} with $B=A$ since it has the simplest form.

System~\eqref{potsysBAspec} also admits only one linearly independent local conservation law~$\mathcal F$ 
which contains the conserved vector $(F,G)=(e^x(e^x+\varepsilon)^{-2}v^3,-e^x(e^x+\varepsilon)^{-1}\int\!A)$ 
associated with the characteristic $(\alpha,\beta)=(e^x(e^x+\varepsilon)^{-2},0)$ and, hence, is induced by a conservation law of 
the initial equation in view of Theorem~\ref{TheoremOnCLsOf2DPotSystems}. 
We choose the solution $\Phi=-(e^x+\varepsilon)^{-1}$ of the equations $D_x\Phi=\alpha$ and $D_t\Phi=-\beta$ and put 
\begin{gather*}
\hat F=\frac{e^xv^3}{(e^x+\varepsilon)^2}-\frac{v^3_x-(e^x+\varepsilon)u}{e^x+\varepsilon}
=-\left(\frac{v^3}{e^x+\varepsilon}\right)_x+u, \\[1ex] 
\hat G=\frac{e^x\int\!A}{e^x+\varepsilon}+\frac{v^3_t-(e^x+\varepsilon)Au_x-\varepsilon \int\!A}{e^x+\varepsilon}
=\left(\frac{v^3}{e^x+\varepsilon}\right)_t-Au_x-\textstyle\int\!A, 
\end{gather*}
Again the conserved vector $(\hat F,\hat G)$ is equivalent to~$(F,G)$ and up to a null divergence
is also equivalent to the conserved vector $(\tilde F,\tilde G)=(u,-Au_x-\textstyle\int\!A)$ 
which depends only on derivatives of~$u$ and belongs to the conservation law $\mathcal F^0$. 
Therefore the ``second-level'' potential system 
\[
v^3_x=(e^x+\varepsilon)u,\quad w^3_x=\frac{e^x}{(e^x+\varepsilon)^2}v^3,
\quad w^3_t=\frac{e^x}{e^x+\varepsilon}\textstyle\int\!A.  
\]
obtained from~\eqref{potsysBAspec} by introducing the ``second-level'' potential $w^3$ with the conservation law~$\mathcal F$ 
is also equivalent, with respect to the point transformation $w^3=v^1-(e^x+\varepsilon)^{-1}v^3$, 
to the united system~\eqref{potsysBAgen}--\eqref{potsysBAspec}. 

The space of conservation laws of the united system~\eqref{potsysBAgen}--\eqref{potsysBAspec} is zero-dimensional. 
Therefore, for any equation~\eqref{eqf1} with $B=A$ all potential conservation laws are induced by local ones 
and all inequivalent potential systems are exhausted by systems~\eqref{potsysBAgen}, \eqref{potsysBAspec} and~\eqref{potsys2BAgen}. 

\medskip

{\noindent\mathversion{bold}$B=\int\! A+uA$.} 
From the point of view of local conservation laws, this case does not differ from the general one. 
Any equation from class~\eqref{eqf1} with such a value of~$B$ and an arbitrary value of~$A$
has the unique linearly independent local conservation law~$\mathcal F^0=\mathcal F^0(A,\int\! A+uA)$. 
At the same time, the corresponding potential system 
\begin{equation}\label{potsysBintAuAgen}\textstyle
v^1_x=u,\quad v^1_t=Au_x+u\int\! A
\end{equation}
also admits the unique linearly independent local conservation law~$\mathcal F^4=\mathcal F^4(A)$ 
with the conserved vector~$(F,G)=(e^{v^1},-e^{v^1}\int\! A)$ and the characteristic 
$(\alpha,\beta)=(e^{v^1},-e^{v^1}\int\! A)$. 
Since the characteristic is completely reduced and depends on the potential~$v^1$, 
in view of Proposition~\ref{PropositionOnEssentialChars}
the conservation law~$\mathcal F^4$ is not induced by a local conservation law of 
the initial equation, i.e., it is a purely potential conservation law. 
The potential system~\eqref{potsysBintAuAgen} is reduced to the potential system~\eqref{potsysBAgen}
by means of the potential hodograph transformation
\begin{equation}\label{HodographXV}
\tilde t=t, \quad \tilde x=v^1, \quad \tilde v^1=x, \quad \tilde u=u^{-1}, \quad \tilde A=u^{-2}A,
\end{equation}
and the conservation law~$\mathcal F^4$ is mapped to the one induced by~$\mathcal F^2$.
The same transformation extended  by the formula $\tilde w=-w+v^1e^x$ to the second-level potential~$w$ introduced with~$\mathcal F^4$ 
also reduces the second-level potential system $v_x=u$, $w_x=e^v$, $w_t=e^v\int\! A$ to system~\eqref{potsys2BAgen}.
As a result, although any equation from class~\eqref{eqf1} with $B=\int\! A+uA$ admits a nontrivial potential conservation law, 
this case does not give principally new potential systems. 

\medskip

{\noindent\bf Linear heat equation.} 
The space of local conservation laws of the linear heat equation $u_t=u_{xx}$
is infinite-dimensional and formed by $\mathcal F^4_h$, 
where $h=h(t,x)$ runs through solutions of the backward linear heat equation $h_t+h_{xx}=0$~\cite{Dorodnitsyn&Svirshchevskii1983}.
Fixing an arbitrary $p\in\mathbb{N}$ and
choosing $p$~linearly independent solutions $h^1$, \dots, $h^p$ of
the backward linear heat equation, we obtain $p$~linearly independent conservation laws $\mathcal F^4_{h^1}$, \dots, $\mathcal F^4_{h^p}$.
In view of Theorem~5 of~\cite{Popovych&Ivanova2004CLsOfNDCEs} (see also Lemma~6 of~\cite{Popovych&Kunzinger&Ivanova2008}),
the potentials $v^1$, \dots, $v^p$ introduced for these conservation laws by 
\begin{equation}\label{potsyslinP}
v^s_x=h^s u,\quad v^s_t=h^s u_x-h^s_xu, \quad s=1,\dots,p,
\end{equation}
are independent in the sense of Definition~\ref{DefinitionOfPotentialDependence}.
According to Theorem~8 of~\cite{Popovych&Ivanova2004CLsOfNDCEs} or Theorem~5 of~\cite{Popovych&Kunzinger&Ivanova2008}, 
any local conservation law of system~\eqref{potsyslinP} is induced by a local conservation law of the linear heat equation. 
As a result, the systems of the form~\eqref{potsyslinP} exhaust all possible potential systems of the linear heat equation 
and all potential conservation laws of this equations are induced by local ones. 
 
\medskip

{\noindent\bf Linearizable equations.} 
Up to $G^\sim$-equivalence, class~\eqref{eqf1} contains three linearizable equations. 
These are 
the $u^{-2}$-diffusion equation $u_t=(u^{-2}u_x)_x$~\cite{Bluman&Kumei1980,Storm1951}, 
the related equation $u_t=(u^{-2}u_x)_x+u^{-2}u_x$~\cite{Fokas&Yortsos1982,Strampp1982b}
and the Burgers equation~$u_t=u_{xx}+2uu_x$~\cite{Forsyth1906,Hopf1950,Cole1951}.
These equations are well known to be linearized by nonlocal transformations
(so-called potential equivalence transformations in the class~\eqref{eqf1} \cite{Popovych&Ivanova2005PETs,LisleDissertation})
to the linear heat equation.
While possessing the usual properties concerning local conservation laws, 
they are distinguished from the other diffusion--convection equations of the form~\eqref{eqf1}
by possessing an infinite number of linearly independent purely potential conservation laws.

The $u^{-2}$-diffusion equation {\mathversion{bold}$u_t=(u^{-2}u_x)_x$} admits, as a subcase of the case~$B=0$,
two linearly independent local conservation laws $\mathcal F^0=\mathcal F^0(u^{-2},0)$ and $\mathcal F^1=\mathcal F^1(u^{-2})$.
The potential system constructed by~$\mathcal F^1$ has the form~\eqref{potsysB0spec} with $A=u^{-2}$ 
and possesses the same properties as for general~$A$ (see the case $B=0$). 
The conservation law~$\mathcal F^0$ gives a potential system of the form~\eqref{potsysB0gen} with $A=u^{-2}$, 
whose space of local conservation laws, in contrast to the general value of~$A$, is infinite-dimensional 
and consists of the conservation laws $\mathcal F^5_\sigma$ 
with the conserved vectors $(F,G)=(\sigma,\sigma_vu^{-1})$ and the characteristics $(\alpha,\beta)=(\sigma_v,-\sigma_tu^{-1})$. 
Here the parameter-function $\sigma=\sigma(t,v)$ runs through the solution set of the backward linear heat equation $\sigma_t+\sigma_{vv}=0$ 
and the potential $v^1$ is re-denoted by $v$. 
Since any of the above characteristics is completely reduced and depends on the potential~$v$ in case of $\sigma_{vv}\ne0$ then 
in view of Proposition~\ref{PropositionOnEssentialChars}
each conservation law~$\mathcal F^5_\sigma$ with $\sigma_{vv}\ne0$ is not induced by a local conservation law of 
the initial equation, i.e., it is a purely potential conservation law. 
The conservation law~$\mathcal F^5_v$ is induced by $\mathcal F^2=\mathcal F^2(u^{-2})$ and  
$\mathcal F^5_1$ is the zero conservation law.

The $u^{-2}$-diffusion equation is reduced to the linear heat equation~\cite{Bluman&Kumei1980} 
by the potential hodograph transformation~\eqref{HodographXV}.
More precisely, the transformation~\eqref{HodographXV} is a local transformation between
the corresponding potential systems $v_x=u$, $v_t=u^{-2}u_x$ and $v_x=u$, $v_t=u_x$
constructed by means of the conservation laws~$\mathcal F^0(u^{-2},0)$ and $\mathcal F^0(1,0)=\mathcal F^4_1$, respectively.
Hence the action of~\eqref{HodographXV} maps each of these conservation laws to zero of the target system.
Moreover, the transformation~\eqref{HodographXV} provides the correspondence between the 
conservation laws $\mathcal F^5_\sigma$ and $\mathcal F^4_h$ with the same values of the parameter-functions $\sigma(t,v)=h(\tilde t,\tilde x)$. 

After fixing an arbitrary $p\in\mathbb{N}$ and choosing $p$~solutions $\sigma^1$, \dots, $\sigma^p$ of
the backward linear heat equation any of whose linear combinations is not a constant, 
we construct the second-level potential system~$\mathcal S$ from system~\eqref{potsysB0gen} with $A=u^{-2}$
using the $p$~linearly independent conservation laws $\mathcal F^5_{\sigma^1}$, \dots, $\mathcal F^5_{\sigma^p}$. 
The system~$\mathcal S$ is pointwise equivalent to the potential system of 
the linear heat equation, associated with the conservation laws $\mathcal F^4_1$, $\mathcal F^4_{\sigma^1}$, \dots, $\mathcal F^4_{\sigma^p}$.
The above results on conservation laws of the linear heat equation imply that 
any conservation law of $\mathcal S$ is induced by a conservation law of system~\eqref{potsysB0gen} with $A=u^{-2}$. 
Consequently, this case does not give principally new potential systems 
although the $u^{-2}$-diffusion equation admits an infinite-dimensional space of first-level potential conservation laws 
connected with system~\eqref{potsysB0gen}.  

Since the equation {\mathversion{bold}$u_t=(u^{-2}u_x)_x+u^{-2}u_x$}
is reduced to the $u^{-2}$-diffusion equation by the point transformation $\tilde t=t$, $\tilde x=e^x$, $\tilde u=e^{-x}u$,
its conservation laws are connected with ones of the linear heat equation in a way similar to the previous case. 
Thus, the space of local conservation laws of the equation $u_t=(u^{-2}u_x)_x+u^{-2}u_x$ is the usual one for the case $B=A$. 
It is generated by two linearly independent conservation laws $\mathcal F^0=\mathcal F^0(u^{-2},u^{-2})$ and $\mathcal F^2=\mathcal F^2(u^{-2})$.
The potential system associated with $\mathcal F^2+\varepsilon\mathcal F^0$ is of the form~\eqref{potsysBAspec} with $A=u^{-2}$. 
Its properties are as usual for the case $B=A$. 
At the same time, the other inequivalent potential system which is associated with $\mathcal F^0$ 
possesses an infinite-dimensional space of local conservation laws, equal to $\{\mathcal F^6_\sigma\}$. 
Here $\mathcal F^6_\sigma$ is a conservation law with the conserved vector $(\sigma e^x,\sigma_vu^{-1}e^x)$ 
and the characteristic $(\sigma_ve^x,-\sigma_tu^{-1}e^x)$. 
The parameter-function $\sigma=\sigma(t,v)$ again runs through the solution set of the backward linear heat equation $\sigma_t+\sigma_{vv}=0$ 
and the potential $v^1$ is re-denoted by $v$. 
Since any of the above characteristics is completely reduced and depends on the potential~$v$ in case of $\sigma_{vv}\ne0$ then 
in view of Proposition~\ref{PropositionOnEssentialChars}
each conservation law~$\mathcal F^6_\sigma$ with $\sigma_{vv}\ne0$ is not induced by a local conservation law of the initial equation, 
i.e., it is a purely potential conservation law. 
At the same time, these conservation laws lead to potential systems which are equivalent to potential systems of the linear heat equation, 
which have form~\eqref{potsyslinP}. 

The Burgers equation~{\mathversion{bold}$u_t=u_{xx}+2uu_x$} is distinguished from the equations of the form~\eqref{eqf1} with $B=\int\! A+uA$ 
through its potential conservation laws. 
As any equation with  $B=\int\! A+uA$, it possesses 
the unique linearly independent local conservation law~$\mathcal F^0=\mathcal F^0(1,2u)$.
The associated potential system $v_x=u$, $v_t=u_x+u^2$ has the infinite-dimensional space of 
conservation laws $\mathcal F^7_h$ with the conserved vectors $(he^v,h_xe^v-hue^v)$ and the characteristics $(he^v,h_xe^v-hue^v)$.  
Here the parameter-function $h=h(t,x)$ runs through the solution set of the backward linear heat equation $h_t+h_{xx}=0$. 
Any of the above characteristics is completely reduced and depends on the potential~$v$ if $h\ne0$. 
Hence in view of Proposition~\ref{PropositionOnEssentialChars}
each conservation law~$\mathcal F^7_h$ with $h\ne0$ is not induced by a local conservation law of the initial equation, 
i.e., it is a purely potential conservation law. 

The potential system $v_x=u$, $v_t=u_x+u^2$ of the Burgers equation~$u_t=u_{xx}+2uu_x$
is mapped to the potential system $\tilde v_{\tilde x}=\tilde u$, $\tilde v_{\tilde t}=\tilde u_{\tilde x}$
(constructed from the linear heat equation~$\tilde u_{\tilde t}=\tilde u_{\tilde x\tilde x}$ 
with the ``common'' conservation law $\mathcal F^0(1,0)=\mathcal F^4_1$)
by the point transformation
\[
\tilde t=t,\quad \tilde x=x,\quad \tilde u=u e^v,\quad \tilde v=e^v.
\]
This transformation establishes the correspondence between the conservation law $\mathcal F^7_{h_x}$ and 
the conservation law of the potential system $\tilde v_{\tilde x}=\tilde u$, $\tilde v_{\tilde t}=\tilde u_{\tilde x}$, 
induced by $\mathcal F^4_h$.
Note that if the parameter-function $h=h(t,x)$ is a solution of the backward linear heat equation 
then its derivative $h_x$ also is a solution of the same equation.
The famous Cole--Hopf transformation \cite{Cole1951,Hopf1950}
(first found in~\cite{Forsyth1906}) is a consequence of the above transformation and in fact linearizes the Burgers equation
to the linear heat equation with respect to the potential~$\tilde v$~\cite{LisleDissertation,Popovych&Ivanova2005PETs}.

For some $p\in\mathbb{N}$ we choose $p$~solutions $h^1$, \dots, $h^p$ of the backward linear heat equation such that 
any of their linear combinations is not a constant. 
The second-level potential system $\mathcal S$ constructed from the potential system $v_x=u$, $v_t=u_x+u^2$ using 
the $p$~linearly independent conservation laws $\mathcal F^7_{h^1_x}$, \dots, $\mathcal F^7_{h^p_x}$ 
is pointwise equivalent to the potential system of the linear heat equation, associated with the conservation laws 
$\smash{\mathcal F^4_1}$, $\smash{\mathcal F^4_{h^1}}$, \dots, $\smash{\mathcal F^4_{h^p}}$.
The above results on conservation laws of the linear heat equation imply that 
any conservation law of $\mathcal S$ is induced by a conservation law of the potential system $v_x=u$, $v_t=u_x+u^2$. 
Therefore this case gives only potential systems which are pointwise equivalent to systems of the form~\eqref{potsyslinP}.

\section{Potential indeterminacy and potential conservation laws}\label{SectionOnPotIndeterminacyAndPotCLs}

Suppose that $\mathcal L_{\rm p}$ is a system determining an Abelian covering of~$\mathcal L$ 
(resp. a potential system of~$\mathcal L$ in the two-dimensional case). 
The potential part of~$\mathcal L_{\rm p}$ consisting of equations~\eqref{EqPotPartInMultiDimCaseForAbelianCoverings} 
defines the potentials $v^1$, \dots, $v^p$ up to arbitrary constant summands. 
This means that the system~$\mathcal L_{\rm p}$ is invariant with respect to the gauge transformations 
of the form $\tilde x_i=x_i$, $\tilde u^a=u^a$ and $\tilde v^s=v^s+c_s$, where $c_s=\const$, 
i.e., the operators $\p_{v^s}$ belong to the maximal Lie invariance group of~$\mathcal L_{\rm p}$. 
It is well known that, acting by an appropriately prolonged generalized symmetry operator of a system of differential equations 
on a conserved vector of the same system, one obtains a conserved vector of this system 
(cf. \cite[Proposition 5.64]{Olver1993}).
Due to the special structure of $\mathcal L_{\rm p}$, the statement on the action by the operators $\p_{v^s}$ 
to conserved vectors of $\mathcal L_{\rm p}$ can be formulated more precisely. 

\begin{proposition}
Any derivative of any conserved vector of~$\mathcal L_{\rm p}$ with respect to potentials is a conserved vector of~$\mathcal L_{\rm p}$. 
The same derivative of a characteristic of the conservation law containing the initial conserved vector 
represents a characteristic associated with the differentiated conserved vector.  
\end{proposition}

\begin{proof}
Let $F\in\CV(\mathcal L_{\rm p})$. 
In view of Proposition~\ref{PropositionOnCharRepresentationOfCVs}, 
there exist differential functions $\bar\lambda^{si}[u,v]$ and $\lambda^\nu[u,v]$ and an $n$-tuple $\hat F$ 
vanishing on the solutions of~$\mathcal L_{\rm p}$ such that  
\[
D_iF^i=\bar\lambda^{si}(v^s_i-G^{si})+\lambda^\nu L^\nu+D_i\hat F^i.
\]
The functions $\bar\lambda^{si}$ and $\lambda^\nu$ are the components of a characteristic of the conservation law containing~$F$. 
For a fixed value of~$s$, we act on the later equality with the infinite prolongation of the operator $\p_{v^s}$, 
which formally coincides with $\p_{v^s}$, and use the property of commutation of any infinitely prolonged operator 
with each total differentiation operator:
\[
D_iF^i_{v^s}=\bar\lambda^{si}_{v^s}(v^s_i-G^{si})+\lambda^\nu_{v^s}L^\nu+D_i\hat F^i_{v^s}.
\]
Since $\p_{v^s}$ is a symmetry operator of~$\mathcal L_{\rm p}$ then $\hat F^i_{v^s}$ vanishes on the solutions of~$\mathcal L_{\rm p}$. 
Therefore, $F_{v^s}$ is a conserved vector of~$\mathcal L_{\rm p}$ and 
$(\bar\lambda^{si}_{v^s},\lambda^\nu_{v^s})$ is a characteristic of the conservation law containing this conserved vector.
\end{proof}

Moreover, there exist an interesting connection between conserved vectors and characteristics 
of the potential systems determining Abelian coverings.

\begin{proposition}\label{PropositionOnCoincidingOfCharsAndCVsOfPotCLs}
For any fixed value of~$s$, the components of an arbitrary characteristic $\lambda$
of a conservation law of the system~$\mathcal L_{\rm p}$, which corresponds to the equations defining the potential~$v^s$, 
form a conserved vector of~$\mathcal L_{\rm p}$ belonging to the conservation law with the characteristic $-\lambda_{v^s}$. 
\end{proposition}

\begin{proof}
Since $\lambda\in\Ch(\mathcal L_{\rm p})$, there exists a conserved vector~$F$ of~$\mathcal L_{\rm p}$ such that 
\begin{equation}\label{EqCharFormOfPotCLs}
\bar\lambda^{si}(v^s_i-G^{si})+\lambda^\nu L^\nu=D_iF^i.
\end{equation}
Applying the component $\Eop_{v^s}$ of the extended Euler operator to equation~\eqref{EqCharFormOfPotCLs}, 
we obtain 
\[
0=D_i\bar\lambda^{si}+D^\alpha(\bar\lambda^{\sigma i}_{v^\sigma_\alpha}(v^\sigma_i-G^{\sigma i}))+D^\alpha(\lambda^\nu_{v^\sigma_\alpha}L^\nu),
\]
where $\alpha$ runs through the multiindex set. 
The derived equality implies that $(\bar\lambda^{s1},\dots,\bar\lambda^{sn})$ is a conserved vector of~$\mathcal L_{\rm p}$ 
since all the summands excluding $D_i\bar\lambda^{si}$ obviously vanish on the solutions of~$\mathcal L_{\rm p}$. 
This equality can be represented as a characteristic form of a conservation law of~$\mathcal L_{\rm p}$:
\[
D_i\bar\lambda^{si}=-\bar\lambda^{\sigma i}_{v^\sigma}(v^\sigma_i-G^{\sigma i})-\lambda^\nu_{v^\sigma}L^\nu
-\sum_{|\alpha|>0}D^\alpha(\bar\lambda^{\sigma i}_{v^\sigma_\alpha}(v^\sigma_i-G^{\sigma i}))
-\sum_{|\alpha|>0}D^\alpha(\lambda^\nu_{v^\sigma_\alpha}L^\nu),
\]
which associates the conserved vector $(\bar\lambda^{s1},\dots,\bar\lambda^{sn})$ with the characteristic
$(-\bar\lambda^{\sigma i}_{v^s},-\lambda^\nu_{v^s})$. 
Therefore, the conserved vector $(\bar\lambda^{s1},\dots,\bar\lambda^{sn})$ is equivalent to the conserved vector $-F_{v^s}$.
\end{proof}

Good illustrative examples for the above statements are given by linearizable convection--diffusion equations 
(cf. the previous section). 
Thus, the potential system $v_x=u$, $v_t=u^{-2}u_x$ of the $u^{-2}$-diffusion equation $u_t=(u^{-2}u_x)_x$ possesses 
the infinite-dimensional space of the local conservation laws $\mathcal F^5_\sigma$ 
with the conserved vectors $(\sigma,\sigma_vu^{-1})$ and the characteristics $(\sigma_v,-\sigma_tu^{-1})$. 
Here the parameter-function $\sigma=\sigma(t,v)$ runs through the solution set of the backward linear heat equation $\sigma_t+\sigma_{vv}=0$. 
Since the derivative $\sigma_v$ also is a solution of the backward linear heat equation, 
the image of the conserved vector $(\sigma,\sigma_vu^{-1})$ under the action of the operator~$\p_v$ 
is a conserved vector belonging to the conservation law $\mathcal F^5_{\sigma_v}$. 
The characteristic $(\sigma_v,-\sigma_tu^{-1})$ of $\mathcal F^5_\sigma$ coincides with this conserved vector from $\mathcal F^5_{\sigma_v}$.
Analogously, the local conservation laws of
the potential system $v_x=u$, $v_t=u^{-2}u_x-u^{-1}$ of the equation $u_t=(u^{-2}u_x)_x+u^{-2}u_x$ is exhausted by 
the conservation laws $\mathcal F^6_\sigma$ with the conserved vectors $(\sigma e^x,\sigma_vu^{-1}e^x)$ 
and the characteristics $(\sigma_ve^x,-\sigma_tu^{-1}e^x)$. 
The action of the operator $\p_v$ maps the conserved vector $(\sigma e^x,\sigma_vu^{-1}e^x)$ to 
the conserved vector $(\sigma_v e^x,\sigma_{vv}u^{-1}e^x)$ which belongs to the conservation law $\mathcal F^6_{\sigma_v}$ 
and coincides with the characteristic $(\sigma_ve^x,-\sigma_tu^{-1}e^x)$ of  $\mathcal F^6_\sigma$.
Any local conservation law of the potential system $v_x=u$, $v_t=u_x+u^2$ associated with the Burgers equation~$u_t=u_{xx}+2uu_x$ 
has a conserved vector and a characteristic of the same form $(he^v,h_xe^v-hue^v)$, 
where the parameter-function $h=h(t,x)$ runs through the solution set of the backward linear heat equation $h_t+h_{xx}=0$. 
The action of~$\p_v$ does not change such conserved vectors and characteristics. 
This explains in view of Proposition~\ref{PropositionOnCoincidingOfCharsAndCVsOfPotCLs} 
why they have the same form. 
The non-linearizable equations admits only potential conservation laws induced by local ones, which are mapped by $\p_v$ to 0.

\subsection*{Acknowledgements}

The authors are grateful to V.~Boyko, G.~Popovych and A.~Sergyeyev for productive and helpful discussions.
The research was supported by START-project Y237 of the Austrian Science Fund.


\begin{thebibliography}{99}
\itemsep=0.ex
\footnotesize
 
\bibitem{Anco&Bluman1997}
Anco~S.C. and Bluman~G., 
Nonlocal symmetries and nonlocal conservation laws of Maxwell's equations, 
{\it J.~Math. Phys.}, 1997, V.38, 3508--3532.

\bibitem{Anco&Bluman2002a}
Anco~S.C. and Bluman~G.,
Direct construction method for conservation laws of partial differential equations.~I.
Examples of conservation law classifications,
{\it Eur. J. Appl. Math.}, 2002, V.13, Part 5, 545--566 (math-ph/0108023).

\bibitem{Anco&Bluman2002b}
Anco~S.C. and Bluman~G.,
Direct construction method for conservation laws of partial differential equations.~II. General treatment,
{\it Eur. J. Appl. Math.}, 2002, V.13, Part 5, 567--585 (math-ph/0108024).

\bibitem{Anco&The2005}
Anco~S.C. and The D., 
Symmetries, Conservation Laws, and Cohomology of Maxwell's Equations Using Potentials,
{\it Acta Appl. Math.}, 2005, V.89, 1--52; arXiv:math-ph/0501052.

\bibitem{Bluman1993}
Bluman G., Potential symmetries and equivalent conservation laws, 
{\it Modern Group Analysis: Advanced Analytical and Computational Methods in Mathematical Physics (Acireale, 1992)}, 
Kluwer Acad. Publ., Dordrecht, 1993, 71--84. 
 
\bibitem{Bluman&Cheviakov&Ivanova2006}
Bluman G., Cheviakov A.F. and Ivanova N.M., 
Framework for nonlocally related partial differential equation systems and nonlocal symmetries: Extension, simplification, and examples,
{\it J. Math. Phys.}, 2006, V.47, 113505,  23 pages.

\bibitem{Bluman&Kumei1980}
Bluman~G. and Kumei~S.,
On the remarkable nonlinear diffusion equation $(\p/\p x)[a(u+b)^{-2}(\p u/\p x)]-\p u/\p t=0$,
{\it J. Math. Phys.}, 1980,  V.21, N~5, 1019--1023.

\bibitem{Blaszak1998}
B\l aczak M.,
Multi-Hamiltonian theory of of dynamical systems, Berlin, Springer, 1998.

\bibitem{Bocharov&Co1997}
Bocharov~A.V., Chetverikov~V.N., Duzhin~S.V., Khor'kova~N.G., Krasil'shchik~I.S.,
Samokhin~A.V., Torkhov~Yu.N., Verbovetsky~A.M. and Vinogradov~A.M.,
{\it Symmetries and conservation laws for differential equations of mathematical physics}, Faktorial, Moscow, 1997.

\bibitem{Cole1951}
Cole J.D.,
On a quasilinear parabolic equation used in aerodinamics,
{\it Quart. Appl. Math.}, 1951, V.9, 225--236.

\bibitem{Dieudonne1972}
Dieudonn{\'e}, J., Treatise on analysis. {V}ol. {III}, 
Academic Press, New York, 1972.

\bibitem{Dorodnitsyn&Svirshchevskii1983}
Dorodnitsyn~V.A. and Svirshchevskii~S.R.,
On Lie--B\"acklund groups admitted by the heat equation with a~source,
Preprint N~101, Moscow,  Keldysh Institute of Applied Mathematics of Academy of Sciences USSR, 1983.

\bibitem{Fokas&Yortsos1982}
Fokas~A.S., Yortsos Y.C.,
On the exactly solvable equation
$S_t = \lbrack (\beta S + \gamma)^{-2}S_x\rbrack_x + \alpha(\beta S + \gamma) ^{-2}S_x$
occurring in two-phase flow in porous media,
{\it SIAM Journal on Applied Mathematics}, 1982, V.42, N~2, 318--332.

\bibitem{Forsyth1906}
Forsyth~A.R.,
The theory of differential equations, V.6, Cambridge, Cambridge University Press, 1906.

\bibitem{Hopf1950}
Hopf E.,
The partial differential equation $u_t+uu_x=\mu u_{xx}$,
{\it Comm. Pure Appl. Math.}, 1950, V.3, 201--230.

\bibitem{Ivanova&Popovych&Sophocleous2004}
Ivanova~N.M., Popovych~R.O. and Sophocleous~C.,
Conservation laws of variable coefficient diffusion--convection equations,
{\it Proceedings of Tenth International Conference in Modern Group Analysis}, (Larnaca, Cyprus, 2004), 107--113.

\bibitem{Krasil'shchik&Vinogradov1984}
Krasil'shchik I.S. and Vinogradov A.M., 
Nonlocal symmetries and the theory of coverings: an addendum to Vinogradov's ``Local symmetries and conservation laws'', 
{\it Acta Appl. Math.}, 1984, V.2, 79--96.

\bibitem{Krasil'shchik&Vinogradov1989}
Krasil'shchik I.S. and Vinogradov A.M., 
Nonlocal trends in the geometry of differential equations: symmetries, conservation laws, and Backlund transformations, 
{\it Acta Appl. Math.},  1989, V.15 , 161--209.

\bibitem{LisleDissertation}
Lisle~I.G.,
Equivalence transformations for classes of differential equations, thesis,
University of British Columbia, 1992.

\bibitem{Marvan2004}
Marvan M., Reducibility of zero curvature representations with application to recursion operators, 
{\it Acta Appl. Math.}, 2004, V.83, 39--68.

\bibitem{Olver1993}
Olver P., Applications of Lie groups to differential equations,
New-York, Springer-Verlag, 1993.

\bibitem{Popovych&Ivanova2004CLsOfNDCEs}
Popovych R.O. and Ivanova N.M.,
Hierarchy of conservation laws of diffusion--convection equations,
{\it J. Math. Phys.}, 2005, V.46, 043502; arXiv:math-ph/0407008.

\bibitem{Popovych&Ivanova2005PETs}
Popovych R.O. and Ivanova N.M.,
Potential equivalence transformations for nonlinear diffusion--convection equations, 
{\it J. Phys. A: Math. Gen.}, 2005, V.38, 3145--3155 (math-ph/0402066).

\bibitem{Popovych&Kunzinger&Ivanova2008}
Popovych R.O., Kunzinger M. and Ivanova N.M., 
Conservation laws and potential symmetries of linear parabolic equations, 
{\it Acta Appl. Math.}, 2008, V.100, 113--185; arXiv:0706.0443.

\bibitem{Sergyeyev2000}
Sergyeyev~A.,
On recursion operators and nonlocal symmetries of evolution equations,
{\it Proceedings of the Seminar on Differential Geometry, Math. Publications},
Silesial University in Opava, Opava, 2000, V.2, 159--173.

\bibitem{Storm1951}
Storm~M. L.,
Heat conduction in simple metals,
{\it J. Appl. Phys.}, 1951, V.22, 940--951.

\bibitem{Strampp1982b}
Strampp~W.,
B\"acklund transformations for diffusion equations,
{\it Physica D}, 1982, V.6, N~1, 113--118.

\bibitem{Tsujishita1982}
Tsujishita T., 
On variation bicomplexes associated to differential equations, 
{\it Osaka J. Math.}, 1982, V.19, 311--363. 

\bibitem{Vinogradov1984}
Vinogradov A. M., 
The ${\cal C}$-spectral sequence, Lagrangian formalism, and conservation laws. 
I. The linear theory, {\it J. Math. Anal. Appl.}, 1984, V.100, 1--40; 
II. The nonlinear theory, {\it J. Math. Anal. Appl.}, 1984, V.100, 41--129.

\bibitem{Vinogradov&Krasil'shchik1984}
Vinogradov A.M., Krasil'shchik I.S., 
On the theory of nonlocal symmetries of nonlinear partial differential equations, 
{\it Dokl. Akad. Nauk SSSR}, 1984, V.275, 1044--1049.

\bibitem{Wahlquist&Estabrook1975}
Wahlquist~H.D. and Estabrook~F.B.,
Prolongation structures of nonlinear evolution equations,
{\it J. Math. Phys.}, 1975, V.16, N~1, 1--7.

\bibitem{Wolf2002}
Wolf~T.
A comparison of four approaches to the calculation of conservation laws,
{\it Eur. J. Appl. Math.}, 2002, V.13, Part 5, 129--152.

\bibitem{Zharinov1986}
Zharinov~V.V.,
Conservation laws of evolution systems,
{\it Teoret. Mat. Fiz.}, 1986, V.68, N~2, 163--171.

\end{thebibliography}
\end{document}